\PassOptionsToPackage{colorlinks,citecolor=blue}{hyperref}
\documentclass[aps,prd,nofootinbib,showpacs,superscriptaddress,preprint]{revtex4-1}
\pdfoutput=1
\usepackage[utf8]{inputenc}
\usepackage{Baskervaldx}
\usepackage[baskervaldx]{newtxmath}
\usepackage{orcidlink}
\usepackage{slashed}
\usepackage{xcolor}
\usepackage{slashed}
\usepackage{algorithm2e}
\usepackage{subcaption}    
\usepackage[font=small,labelfont=bf]{caption}  
\usepackage{comment}
\usepackage{graphicx}
\usepackage{multirow}
\usepackage{float}
\DeclareUnicodeCharacter{2212}{-}

\begin{document}

\title{Light neutrinos, Dark matter and leptogenesis near electroweak scale and $Z_4$ symmetry}
\author{Kunal Pandey \orcidlink{0009-0006-0098-1517}}
\email{kpandey7007@gmail.com}
\affiliation{Centre for Theoretical Physics,\\ Jamia Millia Islamia (Central University), New Delhi - 110025, India}

\author{Rathin Adhikari \orcidlink{0000-0002-8764-9587}}
\email{rathin@ctp-jamia.res.in}
\affiliation{Centre for Theoretical Physics,\\ Jamia Millia Islamia (Central University), New Delhi - 110025, India}

\begin{abstract}
Considering $Z_4$ symmetry in  Type I seesaw scenario,
one could obtain mass-squared differences of light neutrinos, mixings and $CP$ violating phase within $3 \sigma$ confidence level based on neutrino oscillation data. This is possible with only three independent complex parameters for allowed Yukawa couplings and one real mass parameter for heavy right handed neutrino fields around electroweak scale. After considering only three more real parameters as coming from small soft-symmetry breaking terms, the lightest right handed neutrino could be considered as dark matter candidate via freeze-in mechanism and the other two  heavier right handed neutrinos through their decays, could  generate the baryonic asymmetry of the universe naturally via resonant leptogenesis.

\end{abstract}

\maketitle

\section{Introduction}

Neutrino oscillation data  \cite{11} together with cosmological upper bound on light neutrino masses \cite{Planck:2018vyg,DESI:2024hhd} indicate that neutrinos possess very small but non-zero masses ($\sim 0.1$ eV) and hence the Standard Model (SM) needs to be extended in order to accommodate neutrino masses as well as mixing among different flavors of light neutrinos with not too small Yukawa coupling in comparison to unity. The canonical Type-I seesaw \cite{Schechter:1980gr,Schechter:1981cv,Mohapatra:1979ia,Yanagida:1979as,Gell-Mann:1979vob} mechanism remains to this date the most minimal approach of addressing  the mass problem by introducing three heavy Right-handed neutrinos (RHNs) to the particle content of the SM. However, if one considers the neutrino-Higgs Yukawa couplings to be of the order of tau-Higgs Yukawa coupling then the mass scale of the heavy RHNs (also called the seesaw scale) is found to be $\sim 10^9$ GeV. But then, the natural question arises that why the new physics scale is significantly higher than the SM scale. Also, the experimental verification of such a high mass-scale seems difficult in the present experimental context. Is it possible to lower down such a high scale of seesaw mechanism ? To address this issue, various studies in several directions have been performed \cite{Ma:2000cc, Ma:2006km,Haba:2010zi,Kumericki:2012bh,Ker,Adhikari:2010yt,Buchmuller:1990du,Buchmuller:1991tu,Pilaftsis:1991ug} in the context of Type-1 seesaw mechanism. In one category of such works,  the seesaw mass matrix texture has been considered \cite{Buchmuller:1990du,Buchmuller:1991tu,Pilaftsis:1991ug,Ker,Adhikari:2010yt} so that at tree level all light neutrinos are massless but those become massive after considering one loop corrections to the seesaw mass matrix. In another category, which also has considered massless texture, however, apart from three right handed heavy neutrino fields,  additional fields \cite{Buchmuller:1990du,Buchmuller:1990du,Ker,Adhikari:2010yt} have been considered to get light massive neutrinos.
However, such massless texture, in general, leads to some fine tuning of some parameters in the seesaw mass matrix as discussed in \cite{Ker}. In another category of works, no  massless texture has been considered  at the tree level but heavy right handed neutrinos interact with different scalar fields \cite{Ma:2000cc,Ma:2006km,Kumericki:2012bh},with smaller vev and that gives lower seesaw scale.

 In this work, we are interested in those textures of seesaw mass matrix in which three massless neutrinos are obtained at the tree level without any fine tuning between different elements of $M_R$ and $M_D$. Interestingly, such structures of $M_D$ and $M_R$ could also be motivated by a discrete $Z_4$ symmetry for all the field interactions by appropriate charge assignments and this symmetry remain preserved even after Electroweak symmetry breaking. For SM fields, the existence of a remnant $Z_4$ symmetry was noted in \cite{Ma:2023yxq} where the $Z_4$ charges has been correlated to baryon number $B$, lepton number $L$ and hypercharge $Y$ but right handed neutrino mass term violate such $Z_4$ symmetry due to specific charge assignment to all right handed neutrinos. In our work, however , such correlations of $Z_4$ with $B$,  $L$ and $Y$  is not made and with different $Z_4$ charges for different right handed neutrinos, some of their mass term could be invariant under $Z_4$. In this way, $Z_4$ symmetry could be invoked for Type I seesaw mass matrix which results in texture with three massless light neutrinos and will be discussed later in case-(B) in section II.

In this work, with three right handed neutrino fields in addition to SM fields and considering seesaw scale near to electroweak scale, we have addressed three issues: 1) Satisfying neutrino oscillation data 2) Dark matter 3) Baryonic asymmetry of the universe. There are some earlier works \cite{Asaka:2005pn,Asaka:2005an,Datta:2021elq}
considering all these three issues with only three extra right handed neutrino fields. However, unlike previous works, here $Z_4$ symmetry has been considered for all fields. This gives naturally the seesaw mass matrix which makes all three light neutrinos massless at the tree level without any fine tuning of parameters and light neutrinos satisfy  
neutrino oscillation data at $3 \sigma$ confidence level, after considering one loop corrections to the seesaw mass matrix. So the Yukawa couplings involving right handed neutrinos could be larger and possibility of detection of such heavy neutrinos increases in this scenario. Furthermore, the smallness of dark matter coupling with SM fields,  naturally occurs in presence of small soft $Z_4$ symmetry breaking term and the lightest RHN, $N_1$ could be identified as a feebly-interacting dark matter candidate. Such small symmetry breaking terms also play role  in the quasi-degeneacy of two right handed neutrino masses in resonant leptogenesis in which, however, consideration of effective thermal Higgs mass allows only near resonance over small range of temperature near electroweak scale. As the leptogenesis has been considered near electroweak scale, the $CP$-asymmetry corresponding to different flavors of neutrinos in the final state due to heavy RHN decays and washout corresponding to different flavors of neutrinos has been taken into account separately.

In section-\ref{sec1}, we discuss different textures for $M_D$ and $M_R$ for obtaining three massless light neutrinos at tree level. In subsection A, there are discussions on  certain conditions  on the elements of $M_D$ and $M_R$ for massless texture which  lead to fine tuning and in subsection B, there are discussions in which no fine tuning is required.
In section-\ref{sec2}, we discuss the one-loop corrections to both $M_D$ and $M_{L}$ blocks of the full seesaw mass matrix for the seesaw texture as discussed in subsection B of Section II. It has also been discussed when one loop corrections in $M_D$  dominates over corrections in $M_L$ in determining light neutrino masses. In section-\ref{sec5}, we discuss how the lightest RHN  could be a dark matter candidate using freeze-in mechanism and in section-\ref{sec6}, we discuss the possibility of accounting the observed baryonic asymmetry via the Resonant Leptogenesis (RL) mechanism through the decays of other two heavy RHNs. We discuss in brief, the possible search for such heavy neutrinos. Finally, we present our concluding remarks in section-\ref{sec7}.

\section{Massless Texture of Seesaw Mass Matrix At Tree Level :} \label{sec1}

The standard Type-1 seesaw Lagrangian is considered which requires the addition of only three heavy right-handed Majorana neutrinos $(N_R)$ to the Standard Model particle content along with a bare mass term for the $N_R$ fields:

\begin{equation}
                        \mathcal{L}_Y \supset -Y_{\imath \jmath}^{\ell} \Bar{L_\imath}\Phi l_{R\jmath} - Y_{\imath \jmath} \Bar{L}_\imath \widetilde{\Phi} N_{R_{\jmath}}  -\frac{1}{2}\Bar{N}^c_{R_{\imath}} M_{R_{\imath \jmath}} N_{R\jmath}+  h.c
                    \end{equation}
where, we have used $\widetilde{\Phi}= -i \tau_2 \Phi^{*}$. Also, $N_{R_{j}}$ are the three heavy Right-handed neutrinos, $l_{R_{j}}$ are the charged lepton singlets (with $i$ and $j$ indices going from 1 to 3), $L_i$ are the lepton doublets, $\Phi$ is the Higgs-doublet and are given by
\begin{equation}
L_i =
\begin{pmatrix}
\nu_{L_{i}} \\
\ell_{L_{i}}
\end{pmatrix},
\qquad
\Phi =
\begin{pmatrix}
\phi^- \\
\dfrac{1}{\sqrt{2}}\left( v + h + i\phi^3 \right)
\end{pmatrix},
\end{equation}
 where $v=246$ GeV is the \textit{vev} (vacuum expectation value) of the Higgs doublet and $\phi^3$, $\phi^{\pm}$ are the ghost fields. After spontaneous symmetry breaking, the above interaction lagrangian could be written as:

\begin{equation}
-\mathcal{L}_Y \supset Y_{\imath \jmath}^{\ell} \Bar{L_\imath}\Phi l_{R\jmath}+ \frac{1}{2} 
\begin{pmatrix}
\bar{\nu_L} & \bar{N_R^{c}}
\end{pmatrix}
\mathcal{M_{\textit{seesaw}}}
\begin{pmatrix}
\nu_L^c \\
N_R
\end{pmatrix}
+ \text{H.c.},
\label{eq:Lnu}
\end{equation}
where $\mathcal{M_{\textit{seesaw}}}$ is a $6\times6$ matrix and denotes the full Type-I seesaw mass matrix. It consists of four $3\times3$ sub-matrices: $M_{D}$, $M_{R}$, $M^{T}_{D}$ and $M_L$ .Considering the basis $(\nu_e, \nu_{\mu}, \nu_{\tau}, N_{R_{1}}, N_{R_{2}}, N_{R_{3}})$ one can write: 
\begin{equation} \label{eq2}
    \mathcal{M}_{\textit{}{seesaw}}= \begin{pmatrix}
        M_L= 0 & M_{D} \\
        M_{D}^{T} & M_{R}
    \end{pmatrix}
\end{equation}
where $M_L$ consists of bare Majorana masses for the light neutrinos \cite{11}, which, for the Type-1 seesaw setup is a $3 \times 3$ zero matrix. $M_D$ denotes the Dirac mass matrix for the neutrinos and its elements are given by: $M_{D_{ij}} \sim Y_{ij} v$, where $v$ is the Higgs \textit{vev} and $Y_{ij}$ are the Yukawa couplings (see Eq (1)). $M_R$ denotes the mass matrix for the heavy right handed neutrinos and comes from the bare mass term present in the lagrangian. In order to work with massive neutrinos one has to diagonalize the mass matrix. So, using a $(3+k) \times (3+k)$ (where k goes from 1 to 3) unitary matrix $U$, one can write:

\begin{equation}
\mathcal{M}_{\text{diag}} = U^T \mathcal{M_{\textit{seesaw}}} U,
\label{eq:diag}
\end{equation}
and hence, massive neutrinos can be obtained in the mass eigenstate which can be defined by utilizing the diagonalizing matrix $U$
as:

\begin{equation}
\begin{pmatrix}
\nu_L \\
N_R^c
\end{pmatrix}
= U\,P_L\, n
\equiv U\,P_L
\begin{pmatrix}
\nu_1 \\
\nu_2 \\
\nu_3 \\
N_1 \\
\vdots \\
N_k
\end{pmatrix},
\end{equation}where the vector $n$ denotes all the neutrino mass eigenstates and $\nu_{i}$ denotes the three light neutrino mass eigenstates ($i$ goes from 1 to 3) while the three heavy states in mass basis are denoted by $N_{j}$ ($j$ goes from 1 to $k=3$) and $P_{L} = (1-\gamma_5)/2$ is the left-handed projector. So, just like the seesaw mass-matrix $U$ is also a $6 \times 6$ matrix but, one can express the $U$ matrix in a simpler form by expanding in terms of $M_D M_R^{-1}$ ($\sim\Theta$) as:
\begin{equation}
U =
\begin{pmatrix}
U_{\nu\nu} & U_{\nu N} \\
U_{N\nu} & U_{NN}
\end{pmatrix}.
\label{eq:Ublocks}
\end{equation}
where, considering only the leading order, one finds~\cite{Casas:2001sr, Ibarra:2003up}
\begin{align}\nonumber
U_{\nu\nu} &\simeq \left( 1- \frac{\kappa_1}{2} \right ) U_{\text{PMNS}}, \\ \nonumber
U_{\nu N} &\simeq M_D^\dagger M_R^{-1}, \nonumber \\
U_{N\nu} &\simeq -M_R^{-1} M_D U_{\nu\nu}, \nonumber \\  
U_{NN} &\simeq \left( 1- \frac{\kappa_2}{2}\right)I .
\end{align}
where $\kappa_1$ and $\kappa_2$ are at order $\Theta^2$ and are small numbers, $U_{\rm PMNS}$ is the PMNS (Pontecorvo-Maki-Nakagawa-Sakata) matrix \cite{Zyla:2020zbs,Fogli:2006yq} corresponding to the leptonic mixing matrix and $I$ is the $3 \times 3$ identity matrix. Utilizing Eq (5), Eq (7) and Eq (8) one can quantify the mixing between different components of the seesaw setup. Consequently, the light neutrino mass matrix is given by a simple formula:
\begin{eqnarray}
    m_{\nu} \simeq -M_{D} M_{R}^{-1} M_{D}^{T} = U_{\rm PMNS}\, \hat{m}_\nu \, U_{\rm PMNS}^T
\end{eqnarray}
where $m_{\nu}$ is the $3 \times 3$ light neutrino mass matrix while $\hat{m}_\nu = \mathrm{diag}(m_{\nu_1}, m_{\nu_2}, m_{\nu_3})$ is its corresponding  diagonalized matrix with $m_{\nu_i}$ as the light neutrino masses. These light mass eigenvalues are suppressed by the scale of the heavy right handed neutrino mass which is the seesaw scale. The most general structures for $M_{D}$ and the diagonal $M_{R}$ could be written as:

\begin{equation} \label{eq3}
                    M_D=\begin{pmatrix}
                        \lambda_1  & \lambda_2  & \lambda_3 \\
                        \alpha_1 \lambda_1  & \alpha_2 \lambda_2  & \alpha_3 \lambda_3 \\
                        \beta_1 \lambda_1  & \beta_2 \lambda_2  & \beta_3 \lambda_3 
                    \end{pmatrix}
                \end{equation}
                \begin{equation}
                    M_R=\begin{pmatrix}
                        M_1 & M_4 & M_5 \\
                        M_4 & M_2 & M_6\\
                        M_5 & M_6 & M_3
                        \end{pmatrix}
                    \end{equation}
where the elements $\lambda_i$, $\alpha_{i}$, $\beta_{i}$ and $M_{i}$ in general, may be complex. In the subsections below, we discuss two different scenarios of a `massless texture' at the tree level. These textures have a unique property that they lead to all the three light neutrinos to be massless at the tree level itself with a non-vanishing $M_D$ matrix. A non-vanishing $M_D$ matrix means the heavy neutrinos have non-zero interactions with higgs and light neutrinos and could be very interesting from a phenomenological point of view if the corresponding couplings are not too small and the mass scale of heavy right handed neutrinos are not quite high. This paves the road for the usefulness of these massless textures at the tree level if, after one-loop corrections, they could satisfy neutrino masses and mixings obtained from neutrino oscillation data, then this will correspond to considerably larger couplings and smaller mass scale of heavy Right handed neutrinos. There could be quite different scenarios to obtain different kinds of massless textures which belong to two special cases : (I) massless textures with fine-tuning of parameters and (II) massless textures without any fine tuning of parameters, present in the seesaw mass matrix. The fine tuned case is already well studied in the literature but for completeness we mention it in brief.

\subsection{Case-I}

We first discuss the constraints on the seesaw mass parameters in getting three massless light neutrinos. Later on, we will discuss where constraints on the parameters are not required in obtaining massless texture. As shown in  \cite{Ker,Adhikari:2010yt}, at the leading order, for three light neutrinos to be massless, the following conditions are required : 
\begin{eqnarray}
\alpha_1 = \alpha_2 = \alpha_3 \;,
  \nonumber \\
\beta_1 = \beta_2 = \beta_3 \; ,
\end{eqnarray}
in $M_{D}$, then we will have two massless neutrinos at the tree level. This amounts to making any two columns of $M_D$ proportional to each other and on the top of this if one further puts the condition \cite{Adhikari:2010yt}:

\begin{align}
(M_2 M_3 - M_6^2) \lambda_1^2 + (M_1 M_3 - M_5^2) \lambda_2^2 + (M_1 M_2 - M_4^2) \lambda_3^2 + \nonumber  \\
2 (M_6 M_5 - M_3 M_4) \lambda_1 \lambda_2 + 2 (M_4 M_5 - M_1 M_6) \lambda_2 \lambda_3 + 2 (M_4 M_6 - M_2 M_5) \lambda_1 \lambda_3 = 0
\end{align}
then one obtains all the three light neutrinos to be massless at the tree level itself. Such massless-ness for light neutrinos has been extensively studied in the literature \cite{Ker,Buchmuller:1991tu,Buchmuller:1990du,Adhikari:2010yt}. Several variations in the massless-ness conditions for three light neutrinos exist. As for example, instead of Eq.(12) one may consider: 
\begin{equation}
    \lambda_i=0 \quad ,\quad \lambda_j=0 \quad (i\neq j)     
\end{equation}
 together with Eq.(13) to obtain massless-ness. In general, the condition in Eq.(13) leads to some fine tuning of the parameters \cite{Ker} like the Yukawa couplings and right handed neutrino masses. However, there are some special cases for which fine tuning of such parameters may not be required and is discussed in the next sub-section. 

\subsection{Case-II}
 
 If one considers\footnote{Instead of ($i$, $j$) pair chosen as $(1,2)$ in Eq.(14) one could have considered $(1,3)$ or $(2,3)$ which will give different conditions than that in Eq. (15)} $\lambda_1 = \lambda_2 = 0$ and $\lambda_3 \neq 0$ then from Eq. (13) one obtains a condition $M_4 = \pm \sqrt{M_1 M_2}$ ,  using which one may write  $M_D$ and $M_R$ matrices as:
\begin{align}
&M_D =
\begin{pmatrix}
0 & 0 &  \lambda_3 \\
0 & 0 & \alpha_3 \lambda_3 \\
0 & 0 & \beta_3 \lambda_3
\end{pmatrix}
\quad,
\quad
M_R =
\begin{pmatrix}
M_1 &  \pm \sqrt{M_1 M_2} & M_5 \\
 \pm \sqrt{M_1 M_2} & M_2 & M_6 \\
M_5 & M_6 & M_3
\end{pmatrix}\;;
\label{eq:MDa}
\end{align}

\begin{equation}
    M_6 \sqrt{M_1} \neq \pm M_5 \sqrt{M_2}
\end{equation}
where Eq.(16) is required if one wants to use the seesaw formula to obtain light neutrino masses. The structure in Eq. (15) was earlier obtained in \cite{Adhikari:2010yt} in which  $M_4$ in Eq. (11) is constrained to be   $\pm \sqrt{M_1 M_2}$.

At this point it is interesting to note that if one chooses $M_2$ to be zero in Eq.(15) which also amounts to considering $M_4=0$ in Eq. (11),  then there is no constraint from Eq. (13) on the remaining parameters without contradicting Eq. (15 b). This interesting observation leads to one possible form for $M_D$ and $M_R$ matrices, without any fine tuning of the parameters and is given as:

\begin{equation}\begin{array}{cc} \label{eq:MD}
M_D =
\begin{pmatrix}
0 & 0 & k v \\
0 & 0 & \alpha v \\
0 & 0 & \beta v
\end{pmatrix}
\quad,
\quad\quad
M_R =
\begin{pmatrix}
M_1 & 0 & M_5 \\
0 & 0 & M_6 \\
M_5 & M_6 & M_3
\end{pmatrix}
\end{array}
\end{equation}
where, we have replaced $\lambda_3$ as $k v$, $\alpha_3 \lambda_3$ as $\alpha v$ and $\beta_3 \lambda_3$ as $\beta v$ in which $v$ is the Higgs \textit{vev} and $k$, $\alpha$ and $\beta$ are the Yukawa couplings ($Y_{13}$, $Y_{23}$ and $Y_{33}$ respectively, as in Eq (1)). For the above texture, for any arbitrary non-zero finite values of any of the parameters, Eq. (17) always leads to three massless light neutrinos at the tree level.

\begin{table}[h]
\centering
\renewcommand{\arraystretch}{1.3}
\begin{tabular}{|c|c|c|c|c|c|c|c|c|c|c|c|c|c|}
\hline
Particle
& gluon
& $W^\pm$
& $Z$
& $A$
& $Q$
& $u_R$
& $d_R$
& $L$
& $e_R$
& $(\phi^+,\phi^0)$
& $N_1$
& $N_2$
& $N_3$ \\
\hline
$Z_4$ 
& $1$
& $1$
& $1$
& $1$
& $i$
& $-i$
& $-i$
& $i$
& $-i$
& $-1$
& $1$
& $i$
& $-i$ \\
\hline
\end{tabular}
\caption{$Z_4$ realization for the texture of Eq.~(17) with $M_3=M_5=0$ .}
\label{table1}
\end{table}

 It is interesting to note that such texture in Eq (17) with $M_3 = M_5 =0 $ (which neither violate Eq. (16) nor violate the massless-ness requirement at the tree level) could be motivated from symmetry principles. If one considers a discrete $Z_4$ symmetry realization as given in Table-I then such texture is naturally obtained from Eq (1). For the SM fields in Table -I the corresponding $Z_4$ transformations are the same as obtained in \cite{Ma:2023yxq}, however in it, the $Z_4$ transformation of a field is linked to baryon number ($B$), lepton number ($L$) and hypercharge ($Y$) of the corresponding field and same $Z_4$ transformations to all $N_R$ fields, is considered. Here, we do not consider such link.  Another point is that, with the same charge assignment (like 3 for transforming as  $(-i) N_R$ as in \cite{Ma:2023yxq}) to all three $N_R$ fields, the second term in Eq. (1) could be completely allowed but the third term - the mass term for three $N_R$ fields, will be completely $Z_4$ violating. In any case, considering any specific $Z_4$ charge for all $N_R$ fields,  the texture in Eq. (17) for both $M_D$ and $M_R$ can never be obtained. 

 In this work,  three different $Z_4$ charges have been considered for three different right handed neutrinos. Because of this, some interaction terms in the Lagrangian in Eq. (1) are not allowed. The first term with only SM fields are completely allowed. But in the second term , only $Y_{i3}$ is allowed while $Y_{i1}=Y_{i2}=0$. In the third term, only $i=j=1$ case and $i=2, j=3$ are allowed which implies the presence of $M_1$ and $M_6$ in $M_R$ in Eq.(17) with $Z_4$ symmetry. So imposing $Z_4$ symmetry (with $Z_4$ transformation as in Table - I) in the Lagrangian in Eq. (1) allows texture in Eq. (17) for $M_3=M_5=0$. So $M_3$ and $M_5$ may exist but only as soft symmetry breaking terms. 
 One could have obtained  mass term for  $N_R$ fields after spontaneous symmetry breaking in presence of a heavy singlet scalar field ($\sigma$) which does not transform under $Z_4$.
 
As the parameters $M_3$ and $M_5$ in Eq (17) are  soft $Z_4$ symmetry breaking terms in the lagrangian, they could  naturally attain smaller values \cite{tHooft:1979rat} in comparison to other terms. In sec-\ref{sec5} and sec- \ref{sec6}, they are considered to be of the order of $10^{-10}$ GeV for the consideration of dark matter and resonant leptogenesis and play insignificant role as far as light neutrino masses and mixings are concerned in sec- III. So we ignored consideration of $M_3$ or $M_5$ in sec- III for discussion on neutrino masses, mixings and $CP$ violating phase.  To get massive light neutrinos, we study various one-loop corrections to the seesaw mass matrix in the subsequent section and demonstrate how the consideration of one-loop corrections lead to breaking of these massless textures and result in massive neutrinos.

\section{One-Loop Corrections and Light Neutrino Mass Matrix:} \label{sec2}
\begin{figure}[t!] 
  \centering
  \begin{subfigure}[b]{0.45\textwidth}
    \centering
    \includegraphics[width=\linewidth]{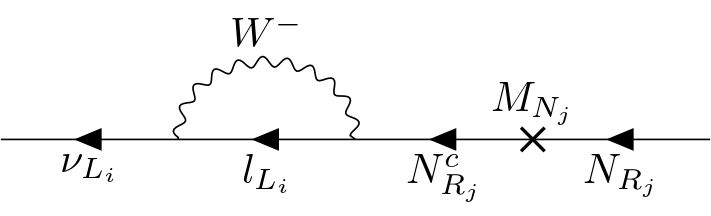}
    \caption{W boson correction}
    \label{fig:subfig1}
  \end{subfigure}
  \hfill
  \begin{subfigure}[b]{0.45\textwidth}
    \centering
    \includegraphics[width=\linewidth]{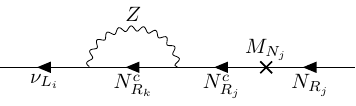}
    \caption{Z boson correction}
    \label{fig:subfig2}
  \end{subfigure}
  
  \vspace{0.3cm}
  \begin{subfigure}[b]{0.45\textwidth}
    \centering
    \includegraphics[width=\linewidth]{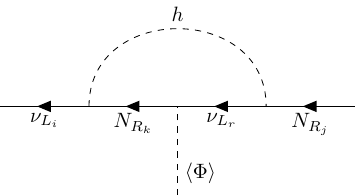}
    \caption{Higgs boson correction (1)}
    \label{fig:subfig3}
  \end{subfigure}
  \hfill
  \begin{subfigure}[b]{0.45\textwidth}
    \centering
    \includegraphics[width=\linewidth]{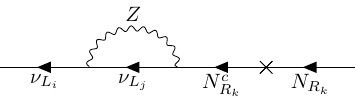}
    \caption{Z boson correction (2)}
    \label{fig:subfig4}
  \end{subfigure}

  \vspace{0.3cm}
  \begin{subfigure}[b]{0.45\textwidth}
    \centering
    \includegraphics[width=\linewidth]{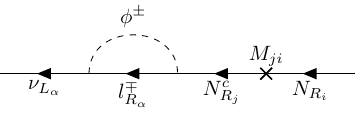}
    \caption{$\phi^\pm$ correction (external insertion)}
    \label{fig:subfig5}
  \end{subfigure}
  \hfill
  \begin{subfigure}[b]{0.45\textwidth}
    \centering
    \includegraphics[width=\linewidth]{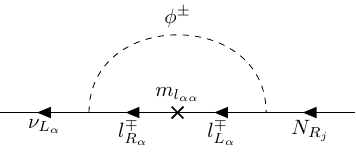}
    \caption{$\phi^\pm$ correction (internal insertion)}
    \label{fig:subfig6}
  \end{subfigure}


  \vspace{0.3cm}
  \begin{subfigure}[b]{0.45\textwidth}
    \centering
    \includegraphics[width=\linewidth]{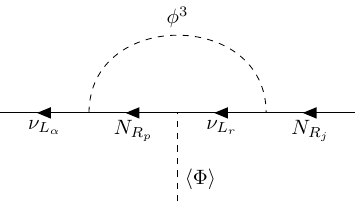}
    \caption{$\phi^3$ correction }
    \label{fig:subfig5}
  \end{subfigure}
 \hfill
  \begin{subfigure}[b]{0.45\textwidth}
    \centering
    \includegraphics[width=\linewidth]{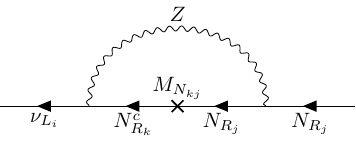}
    \caption{$Z$ boson correction with internal mass insertion }
    \label{fig:subfig6}
  \end{subfigure}

  \caption{Various loop corrections involving gauge and Higgs bosons.}
  \label{fig1}
\end{figure}

In this section, in order to calculate the light neutrino masses, we consider the interplay of various one-loop corrections to different blocks of the seesaw mass matrix. Higher order corrections to massless texture has been studied earlier \cite{Adhikari:2010yt,Ker} to attain appropriate light neutrino masses with a TeV scale of seesaw. However, both these works utilize a massless texture which introduces severe fine-tuning in their setup. Needless to say that in our work there is no fine-tuning and except three heavy right handed neutrino fields there are no other extra scalar fields or any other fields beyond the SM particle content. Working in the Feynman gauge, we have identified all possible one-loop Feynman diagrams involving $N_{Rj}$ and SM fields like $W$, $Z$ and Higgs boson as well as the ghost fields ($\phi^3$ and $\phi^{\pm}$) as shown in Fig \ref{fig1} which contribute directly to $M_D$. There is one crucial difference in the calculation of one-loop corrections (which are self-energy corrections and were calculated using PACKAGE-X \cite{Patel:2015tea,Patel:2016fam}) between $M_D$ and $M_L$ blocks. While evaluating the mass corrections to $M_L$, the $\slashed{p}$ corresponding to external momentum in the one-loop diagrams, should be replaced by zero. However, for $M_D$ mass corrections, such $\slashed{p}$ should be replaced by the appropriate tree level elements of $M_D$ to which the mass corrections has been considered. This is because, all the elements of $M_L$ matrix are zero at the tree level which is not the case for all the $M_D$ matrix elements. 

It should be noted that both the light neutrino Yukawa interaction and the bare mass term for the heavy RHNs in Eq. (1) are of mass dimension four and the theory is renormalizable at one loop level. It should be pointed out that there is mutual cancellation of divergences among some of the diagrams. For instance, the divergences appearing in Fig. (1c) and Fig. (1g) cancel with each other. Also, the diagram in Fig. (d) gives zero contribution to the $M_D$ block. For the remaining diagrams which contain divergences, the corresponding counter terms are available. The one-loop expressions given in the Appendix (\ref{App}) contain the renormalization scale $\mu$, which is a parameter arising from dimensional regularization. For the purpose of numerical evaluation, we make the convenient choice of setting $\mu$ equal to the mass scale of the heavy right-handed neutrino, which is a standard practice in such calculations. 

The diagrams involving $W$, $Z$, $\phi^3$, $\phi^{\pm}$ and $h$ fields are described by the following interactions \cite{Alonso:2012ji} in the mass basis:
\begin{align}
\mathcal{L}^{W^\pm}
&=
\frac{g_W}{\sqrt{2}}\,W^-_\mu\,
\bar{\ell}_\alpha \gamma^\mu U_{\alpha i} P_L n_i
+ \text{h.c.}
\\[1ex]
\mathcal{L}^{Z}
&=
\frac{g_W}{4 c_W}\,Z_\mu\,
\bar{n}_i \gamma^\mu
\left[
C_{ij} P_L - C_{ij}^* P_R
\right]
n_j
\end{align}
\begin{align}
\mathcal{L}^{\phi^\pm}
&=
-\frac{g_W}{\sqrt{2} M_W}\,\phi^- \,
\bar{\ell}_\alpha U_{\alpha i}
\left(
m_{\ell_\alpha} P_L - m_{n_i} P_R
\right)
n_i
+ \text{h.c.}
\\[1ex]
\mathcal{L}^{\phi^3}
&=
-\frac{i g_W}{4 M_W}\,\phi^3\,
\bar{n}_i
\left[
C_{ij}\left(m_{n_i} P_L - m_{n_j} P_R\right)
- C_{ij}^*\left(m_{n_i} P_R - m_{n_j} P_L\right)
\right]
n_j
\\[1ex]
\mathcal{L}^{h}
&=
-\frac{g_W}{4 M_W}\,h\,
\bar{n}_i
\left[
C_{ij}\left(m_{n_i} P_L + m_{n_j} P_R\right)
+ C_{ij}^*\left(m_{n_i} P_R + m_{n_j} P_L\right)
\right]
n_j
\end{align}

\noindent
where $P_{R,L}= (1\pm \gamma^5)/2$, $g_W$ is the gauge coupling constant, $C_W = \cos\theta_{W}$, $\theta_{W}$ is the weak mixing angle, $U$ is the rotation matrix defined earlier in Eq (6). Also, the mass eigenstates of the neutrinos (both light and heavy counterparts) are denoted by $n$ while $l_\alpha$ denotes the mass eigenstate of the charged lepton `$\alpha$', $m_{l_\alpha}$ denotes the mass of the charged lepton `$\alpha$'. In the above equations, $m_{n}$ and $C$ are $(3+k) \times (3+k)$ matrices defined as: 

\begin{equation}
C_{ij} \equiv \sum_{\alpha=1}^{3} U_{i\alpha}^\dagger U_{\alpha j},
\qquad
m_n = \mathrm{Diag}(m_{n_i})
= \mathrm{Diag}\!\left(m_{\nu_1},\, m_{\nu_2},\, m_{\nu_3},\, M_{N_1},\, \ldots,\, M_{N_k}\right).
\end{equation}
where the first three elements in $m_{n}$ correspond to light neutrino masses while the last $k$ (here $k$ goes from $1$ to $3$) entries are for the heavy neutrino masses.

The full expressions of the various one-loop corrections ($\epsilon^{W}$, $\epsilon^{Z}$, $\epsilon^{H}$, $\epsilon^{\phi^3}$ and $\epsilon^{\phi^\pm}$) are given in the Appendix(\ref{App}) in which contributions from all the one loop Feynman diagrams as shown in Fig. (1), have been summed over as:
\begin{equation}
    \epsilon_{ij} = \epsilon_{ij}^{W} + \epsilon_{ij}^{Z} + \epsilon_{ij}^{H} + \epsilon_{ij}^{\phi^3}+ \epsilon_{ij}^{\phi^\pm}
\end{equation}

To study the loop corrections one has to go to the diagonal basis $\widetilde{N_{R}}$ after rotating the non-diagonal basis $N_R$. Considering $M_3$ and $M_5$ in Eq (17) to be zero as discussed earlier, only the 2-3 block of $M_R$ in Eq (17) needs to be diagonalized as follows: 
 
\begin{equation}
                    \begin{pmatrix}\widetilde{N_2}\\
                    \widetilde{N_3}\end{pmatrix} =\begin{pmatrix}
                        \iota\cos\theta  & - \iota\sin \theta \\
                        \sin \theta & \cos \theta
                         \end{pmatrix} 
                         \begin{pmatrix}{N_2}\\
                    {N_3}\end{pmatrix}.
\end{equation}
where $\theta=\frac{\pi}{4}$ and one has:
\begin{align}
    \widetilde{N_2} &= \iota N_2 \cos{\theta} - \iota N_3 \sin{\theta} \\
    \widetilde{N_3} &= N_2 \sin{\theta} + N_3 \cos{\theta}
\end{align}
where the presence of $\iota$ makes sure that both eignevalues come out to be positive. This leads to a diagonal mass matrix $\widetilde{M_R}$ corresponding to $M_R$ and is given as:

\begin{equation}
                    \widetilde{M_{R}}=\begin{pmatrix}
                        M_1  & 0  &  0 \\
                        0  & \widetilde{M_2}  & 0  \\
                        0 & 0  & \widetilde{M_3}   
                    \end{pmatrix}.
\end{equation}
where $\widetilde{M_2} =\widetilde{M_3} = M_6$. Accordingly, the neutrino Dirac mass matrix $M_D$, has to be transformed to $\widetilde{M_D}$ in this new $\widetilde{N_{R}}$ basis as follows:

 \begin{align}
   \widetilde{M_{D_{\alpha 2}}}= v\widetilde{Y_{\alpha 2}} &= v(i Y_{\alpha 2} \cos \theta -i Y_{\alpha 3} \sin\theta) \\ \nonumber
   \widetilde{M_{D_{\alpha 2}}}= v\widetilde{Y_{\alpha 3}} &= v(Y_{\alpha 2} \sin \theta +  Y_{\alpha 3} \cos\theta)
\end{align}
where, Y and $\widetilde{Y}$ are the Yukawa coupling matrix in the basis of $M_R$ and $\widetilde{M_R}$ respectively. Furthermore, as per Eq. (17), $Y_{\alpha 2}=0$, $Y_{1 3}= k$, $Y_{2 3}= \alpha$ and $Y_{3 3}= \beta$. This makes the second column of $\widetilde{M_D}$ non-zero with diagonal $\widetilde{M_R}$ while preserving the massless texture. The corresponding rotated Dirac mass matrix ($\widetilde{M_D}$) for the massless texture is given as:
\begin{equation}
                    \widetilde{M_{D}}=\begin{pmatrix}
                        0   & -\iota  \, k v \sin{\theta}   &  \, k v \cos{\theta}  \\
                        0  & -\iota  \, \alpha v \sin{\theta}   & \, \alpha v \cos{\theta}  \\
                        0 & -\iota  \, \beta v \sin{\theta}   & \, \beta v \cos{\theta}    
                    \end{pmatrix}.
\end{equation}
with $\theta = \frac{\pi}{4}$ as mentioned below Eq. (25). Furthermore, Eq.(28) and Eq.(30) also correspond to massless texture corresponding to Eq.(17) with $M_3=M_5=0$. It should also be noted that $N_1$ is decoupled from $\widetilde{N_2}$ and $\widetilde{N_3}$ and do not interact with SM fields at this stage. 
  
\vspace{1mm}
\noindent

The one-loop corrections involving $W$, $Z$, Higgs boson and ghost fields will modify only the second and third columns of $\widetilde{M_D}$ and the corresponding Feynman diagrams are given in Fig-\ref{fig1}. After this modification one obtains the one-loop corrected Dirac mass matrix $\widetilde{M_D}^{'}$. One may note that, due to the structure of Eq. (30) $\epsilon_{i1}$ is zero. Corresponding to the Eq.(30), one observes the characteristic feature of massless texture that: $\frac{\widetilde{M_{D_{j2}}}}{\widetilde{M_{D_{j3}}}}= -\iota$, where $j=1,2\, \text{and}\,3$. However, the above mentioned loop corrections modify the massless texture as the one-loop corrected ratio: $\frac{\widetilde{M_{D_{j2}}}^{'}}{\widetilde{M_{D_{j3}}}^{'}} \neq -\iota$, for any $j=1,2\, \text{and}\,3$ and also these ratios are different for different values of $j$ in it, as is evident from the various one-loop corrections shown in Appendix-(\ref{App}). These lead to one massless and two massive light neutrinos
 as all the elements in the first column \cite{Adhikari:2010yt} in Eq. (30) is still zero. However, in the context of dark matter discussion in Section IV, after including soft breaking term $M_5$, the first column would be non-zero resulting in non-zero mass for all three light neutrinos, although it would be very small for the lightest one in comparison to the other two. 

Another crucial outcome of the massless texture of Eq (17) is that if one goes on to calculate the light neutrino mass matrix using the seesaw formula Eq (9) then it comes out to be a $3 \times 3$ zero matrix as shown below:
\begin{equation}
    \label{eq22}
  \widetilde{\mathcal{M}_\nu} = - \widetilde{M_D} \widetilde{ M_R}^{-1} \widetilde{M_D}^{T} \equiv 0
\end{equation}
After one loop corrections, $ \widetilde{M_{D}}$ gets modified to $\widetilde{M_{D}}^{'} $ which is just $ \widetilde{M_D} + \epsilon $ where $\epsilon$ is the one-loop correction matrix whose elemets are $\epsilon_{ij}$ as per Eq (24). Furthermore, the one-loop corrections to the $\widetilde{M_R}$ matrix are very small \cite{LEUNG1983461} as compared to the tree level values in $ \widetilde{M_R}$ matrix. Then, one can write the one-loop corrected light neutrino mass matrix as (ignoring higher order terms in $\epsilon$) :

\begin{eqnarray}\label{eq25}
    \widetilde{\mathcal{M}_{\nu}}' \approx -\epsilon \widetilde{M_{R}}^{-1} \widetilde{M_{D}}^{T} - \widetilde{M_{D}} \widetilde{M_{R}}^{-1} \epsilon^{T}            \end{eqnarray}

One may note that $\widetilde{\mathcal{M}_{\nu}}'$ is independent of $M_1$ present in Eq (28) because $\epsilon_{i1}$ as well as $\widetilde{(M_{D})_{i1}}$ elements (where $i=1,2,3$) vanish in our case as discussed earlier. With massless texture of the light neutrino masses at the tree level, Eq.(32) may be considered as the modified seesaw formula for massive light neutrino mass matrix. 
The light neutrino masses could be accommodated at a much lower scale of $\widetilde{M_{R}}$, for instance, for $\epsilon \sim 10^{-7}$ GeV and $\widetilde{M_{D_{ij}}} \sim 0.1$ GeV one can obtain light neutrino masses around 0.1 eV for $\widetilde{M_R}$ around TeV scale or below.

\begin{figure}[h!]
  \centering
  \begin{subfigure}[b]{0.4\textwidth}
    \centering
    \includegraphics[width=\textwidth]{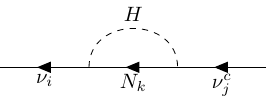}
    \caption{W boson correction}
    \label{fig:subfig1}
  \end{subfigure}
  \hfill
  \begin{subfigure}[b]{0.4\textwidth}
    \centering
    \includegraphics[width=\textwidth]{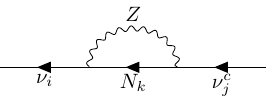}
    \caption{Z boson correction}
    \label{fig:subfig2}
  \end{subfigure}
 \caption{One-loop corrections to the $M_L$}
  \label{fig2}
\end{figure}

However, $M_L$, which is zero at the tree level, becomes non-zero ($\delta M_L$) after one-loop corrections. There are Feynman diagrams as shown in Fig-\ref{fig2} involving Higgs and Z-bosons which give non-zero contributions to the $M_L$ block of seesaw  and they directly affect the light neutrino masses. So, if one accounts for all the loop corrections to $\widetilde{M_D}$ and $M_L$ together, then Eq (4) is modified as:

\begin{equation} \label{eq2}
    \mathcal{M}_{\textit{}{seesaw}}^{(1)}= \begin{pmatrix}
        \delta M_L & \widetilde{M_{D}} + \epsilon \\
        ( \widetilde{M_{D}} + \epsilon)^{T} & \widetilde{M_{R}}
    \end{pmatrix}
\end{equation}
 Following Eq (32), the light neutrino masses are now given as:
 \begin{eqnarray}\label{eq25}
    \widetilde{\mathcal{M}_{\nu}}' \approx \delta \mathcal{M}_{L} -\epsilon \widetilde{M_{R}}^{-1} \widetilde{M_{D}}^{T} - \widetilde{M_{D}} \widetilde{M_{R}}^{-1} \epsilon^{T}            \end{eqnarray}
where the $3 \times 3$ matrix $\delta M_L$ arising from one-loop contributions from Higgs and Z bosons \cite{AristizabalSierra:2011mn} is written as:

\begin{equation} \label{eq:deltaML}
\delta \mathbf{M}_L = \widetilde{\mathbf{M}_D}^T \widetilde{\mathbf{M}}_R^{-1} 
\left\{ 
\frac{g_{W}^{2}}{64 \pi^2 M_W^2} 
\left[
m_h^2 \ln \left( \frac{\widetilde{\mathbf{M}}_R^2}{m_h^2} \right)
+ 3 M_Z^2 \ln \left( \frac{\widetilde{\mathbf{M}}_R^2}{M_Z^2} \right)
\right]
\right\} 
\widetilde{\mathbf{M}_D} 
\end{equation}
Corrections in $\widetilde{\mathcal{M_{\nu}}}'$,  due to one-loop correction ($\epsilon$) in $M_D$, in comparison to $\delta \mathbf{M}_L$, get further suppressed because of suppression factor $\frac{ \widetilde{M_D}}{\widetilde{M_R}}$ in second and third term in Eq (34). However, one-loop corrections $\delta M_L$ to $M_L$ is not multiplied by an such suppression factor in Eq (34). Because of this, in general, the one-loop corrections $\delta M_L$ has more dominating \cite{Grimus:2002nk,AristizabalSierra:2011mn} effect in $\widetilde{\mathcal{M_{\nu}}}'$ than the one-loop corrections in $M_D$. 

However, for our texture of $\widetilde{M_{D}}$ in Eq (30), the first column is zero and because of this $M_1$ in $\widetilde{M_{R}}$ does not play any role in $\delta M_L$. Using Eq (28) and Eq (30) in Eq (35), one finds $\delta M_L$ to be of the following form:
    
\begin{equation}
\delta M_L = 
\underbrace{
\begin{pmatrix}
k^2 & \alpha k & \beta k \\
\alpha k & \alpha^{2} & \alpha \beta \\
\beta k & \alpha \beta & \beta^{2}
\end{pmatrix}
}_{Y_{M_L}}
f(\widetilde{M_{2}}, \widetilde{M_{3}})
\end{equation}

  where the loop function $f(\widetilde{M_{2}}, \widetilde{M_{3}})$ has been factored out and is given by: 
 \begin{align}
f(\widetilde{M_{2}}, \widetilde{M_{3}}) = \, & 
\frac{g_{W}^{2} v^2}{64\, \widetilde{M_{2}} \widetilde{M_{3}} M_W^2 \pi^2} \Bigg[- \widetilde{M_{3}} \sin^2 \theta \left(
M_H^2  \log \left( \frac{\widetilde{M_{2}}^2}{M_H^2} \right)  +
 3  M_Z^2 \log \left( \frac{\widetilde{M_{2}}^2}{M_Z^2} \right) \right)
\nonumber \\
& \quad + \widetilde{M_{2}} \cos^2 \theta  
\left( 
M_H^2 \log \left( \frac{\widetilde{M_{3}}^2}{M_H^2} \right) 
+ 3 M_Z^2 \log \left( \frac{\widetilde{M_{3}}^2}{M_Z^2} \right)
\right)
\Bigg]
\end{align}
Since, in our case $\widetilde{M_2} =\widetilde{M_3} = M_6$ and $\theta=\frac{\pi}{4}$ so, $f(\widetilde{M_{2}}, \widetilde{M_{3}})$ vanishes. Then, it follows from Eq (36) that $\delta M_L$ also vanishes. Hence, in our case, not the loop corrections in $M_L$ but the loop-corrections in  $M_D$ dominate in contributing to $\widetilde{\mathcal{M_{\nu}}}'$. This is in contrast to the earlier general comment regarding the loop corrections to $\widetilde{\mathcal{M_{\nu}}}'$. Hence, for the computation of light neutrino masses, the one-loop corrections only to $\widetilde{M_D}$ as discussed earlier, will be relevant.

\subsection{Light Neutrino masses and mixing and $CP$ violating phase}
In order to compute light neutrino masses, mixing and the $CP$-violating phase, we adopt a numerical approach to diagonalize the one-loop light neutrino mass matrix ($\widetilde{\mathcal{M_{\nu}}}'$) in Eq (34) as $m_{\nu}$ in Eq (9) where the eigenvalues corresponds to the light neutrino masses and the diagonalizing matrix corresponds to the PMNS matrix. In general, after the diagonalization procedure, one should analyze for both, the normal and inverted hierarchy of the mass-squared differences. Consequently, the diagonalizing matrix needs to be properly adjusted based on the eigenvectors for the eigenvalues following specific hierarchy. Then, it is quite straightforward to obtain the three mixing angles and the CP-violating phase as they could be calculated in terms of the elements of the PMNS matrix \cite{Zyla:2020zbs, Fogli:2006yq,Esteban:2020cvm}. The PMNS matrix appears in the weak charged-current interactions of charged leptons and massive neutrinos and is given as:

\begin{equation}
-\mathcal{L}_{\text{cc}}
=
\frac{g}{\sqrt{2}}\,
\begin{pmatrix}
\overline{e_{L}}, & \overline{\mu_{L}}, & \overline{\tau_{L}}
\end{pmatrix}
\,\gamma^\mu\,
U_{\text{PMNS}}
\begin{pmatrix}
\nu_1 \\
\nu_2 \\
\nu_3
\end{pmatrix}_{\!L}
\,W^+_\mu
+ \text{h.c.},
\label{eq:cc}
\end{equation}
where $e$, $\mu$ and $\tau$ are the mass eigenstates for electron, muon and tau respectively. In the basis where the flavor eigenstates of the three charged leptons are identical with their mass eigenstates we have the following relation between the flavor and mass basis of the light neutrinos:

\begin{equation}
\begin{pmatrix}
\nu_e \\
\nu_\mu \\
\nu_\tau
\end{pmatrix}_{\!L}
=
U_{\nu \nu}
\begin{pmatrix}
\nu_1 \\
\nu_2 \\
\nu_3
\end{pmatrix}_{\!L}
\approx
\begin{pmatrix}
U_{e1} & U_{e2} & U_{e3} \\
U_{\mu 1} & U_{\mu 2} & U_{\mu 3} \\
U_{\tau 1} & U_{\tau 2} & U_{\tau 3}
\end{pmatrix}_{\text{PMNS}}
\begin{pmatrix}
\nu_1 \\
\nu_2 \\
\nu_3
\end{pmatrix}_{\!L}
\label{eq:pmns}
\end{equation}
where using Eq.(7) and Eq.(8) and ignoring $\kappa_1$, $U_{\nu \nu}$ has been treated as $U_{\text{PMNS}}$. Considering the standard parametrization of $U_{\text{PMNS}}$ \cite{ParticleDataGroup:2022pth}, one can relate with three mixing angles (the reactor ($\theta_{13}$), solar ($\theta_{12}$) and the atmospheric ($\theta_{23}$)) with elements of $U_{\text{PMNS}}$ as: 

\begin{align}
    \sin \theta_{13} = |U_{e3}|; \qquad   \sin^2 \theta_{12} 
=
\frac{|U_{e2}|^2}{1 - |U_{e3}|^2};   \qquad   \sin^2 \theta_{23}
=
\frac{|U_{\mu 3}|^2}{1 - |U_{e3}|^2}
\end{align}

One can find the $CP$-violating phase $\delta$ in the standard parametrization by using the Jarlskog invariant ($J_{\rm CP}$) quantity \cite{Jarlskog:1985ht,Xing:2020ald} which is defined in terms of the elements of $U$ as follows:

\begin{align}
J_{\rm CP}
= 
\operatorname{Im}
\left[
U_{e1} U_{\mu 2} U_{e2}^\ast U_{\mu 1}^\ast
\right]
\quad
=
c_{12} s_{12}\,
c_{23} s_{23}\,
c_{13}^2 s_{13}\,
\sin\delta
\end{align}
where we have:
\[
c_{ij} \equiv \cos\theta_{ij},
\qquad
s_{ij} \equiv \sin\theta_{ij}.
\]

\begin{table}[h!]
\centering
\begin{tabular}{|c|c|c|c|c|c|c|c|}
\hline
\textbf{Benchmark Point} & $k_{r}$  & $k_{i}$  & $\alpha_{r}$  & $\alpha_{i}$  & $\beta_{r}$  & $\beta_{i}$  & $M_6$ (GeV) \\ \hline
\textbf{BP1} & 0.0000035 & 0.000225 & 0.0004 & 0.0000285 & 0.00035 & 0.00065 & 152 \\ \hline
\textbf{BP2} & 0.0000045 & 0.00025 & 0.00045 & 0.00004 & 0.000375 & 0.000775 & 210 \\ \hline
\textbf{BP3} & 0.000005 & 0.0003 & 0.000525 & 0.00002 & 0.00045 & 0.00085 & 300 \\ \hline
\textbf{BP4} & -0.00004 & 0.0004 & 0.0008 & 0.00002 & 0.0006 & 0.001 & 600 \\ \hline

\end{tabular}
\caption{Parameter values satisfying $3\sigma$ limits for light neutrino mass and mixing} \label{tab1}
\label{tab:bp_parameters}
\end{table}

\begin{table}[h!]
\centering
\renewcommand{\arraystretch}{1.2}
\begin{tabular}{|c|c|c|c|c|c|c|}
\hline
\textbf{Benchmark Point} 
& $ \theta_{12}$ 
& $\theta_{23}$ 
& $ \theta_{13}$ 
& $\delta_{\rm CP}$ 
& $\Delta m^2_{21}\;(\text{eV}^2)$ 
& $\Delta m^2_{31}\;(\text{eV}^2)$ \\ 
\hline

\textbf{BP1}
& $34.93^{\circ}$ 
& $41.19^{\circ}$ 
& $8.56^{\circ}$ 
& $ 313.54^{\circ}$
& $0.0000694 $ 
& $0.00246$ \\ \hline

\textbf{BP2}
& $32.74^{\circ}$ 
& $41.87^{\circ}$ 
& $8.59^{\circ}$ 
& $ 312.26^{\circ}$
& $0.0000803 $ 
& $0.00252$ \\ \hline
\textbf{BP3}
& $35.42^{\circ}$ 
& $41.43^{\circ}$ 
& $8.53^{\circ}$ 
& $ 314.92^{\circ}$
& $0.0000798 $ 
& $0.00250$ \\ \hline

\textbf{BP4}
& $33.72^{\circ}$ 
& $48.56^{\circ}$ 
& $8.24^{\circ}$ 
& $ 299.26^{\circ}$
& $0.0000714$
& $0.00253$ \\ \hline

\end{tabular}
\caption{Neutrino mixing angles, CP-violating phase, and mass-squared differences corresponding to the parameters listed in Table~\ref{tab:bp_parameters} , all of which are within $3\sigma$ experimental limits.}
\label{tab:pmns_observables} 
\end{table}

Using $\widetilde{\mathcal{M}_{\nu}}'$ in Eq (34) as $m_{\nu}$ in Eq (9) one can obtain the $U_{\text{PMNS}}$ following the diagonalization procedure as shown in Eq (9). However, to obtain the three mixing angels and one $CP$-violating phase in $U_{\text{PMNS}}$ and the eigenvalues indicating the mass-squared differences to be within $3 \sigma$ limit \cite{Esteban:2024eli,NuFIT} based on neutrino oscillation experimental data, one needs to scan for the appropriate values of the relevant parameters ($k$, $\alpha$, $\beta$ as complex and $\widetilde{M_2}= \widetilde{M_3}= $ $M_6$ as real in our case) present in the $\widetilde{M_R}$ and $\widetilde{M_D}$ matrices in Eq (28) and Eq (30) respectively. As mentioned earlier, $M_1$ in $\widetilde{M_R}$ does not play any role in elements of $\widetilde{\mathcal{M}_{\nu}}'$. We present benchmark points (BP) in TABLE-\ref{tab:bp_parameters} that correspond to the mass-squared differences, mixing and the CP-violating phase which satisfies the neutrino oscillation experimental data within  $3 \sigma$ limits for the normal hierarchy case \footnote{cosmological observations indicate preference towards the normal hierarchy case \cite{DESI:2024hhd}}. In Table-(II) , the subscript `$r$' and `$i$' correspond to real values and imaginary values respectively of the couplings $k$, $\alpha$ and $\beta$ respectively.
For any particular right handed neutrino mass corresponding to $M_6$, from Table-(II) it is observed that $|k| < |\alpha| < |\beta|$. Also all these parameters are further smaller with the lower values of $M_6$. Although we have shown in Table -II and III,  the masses of two heavier right handed neutrinos somewhat nearer the electroweak scale, however we have also verified that even for $M_6$ nearer to about 950 GeV, it is possible to satisfy experimental data within $3\sigma$ limits. It is remarkable that in presence of $Z_4$ symmetry, with only three complex parameters corresponding to $M_D$ and one real parameter in $M_R$, it is possible to explain the neutrino oscillation data.

\section{ Dark Matter Candidate} \label{sec5}
There is a possibility to identify the decoupled RHN ($N_1$) as a feebly interacting massive particle (FIMP) to act as a dark matter candidate via the freeze-in \cite{Hall:2009bx} mechanism provided that the soft breaking $M_5$ is considered in $\widetilde{M_R}$ in Eq. (28). In our setup, demanding a soft $z_4$ symmetry breaking small off-diagonal element ($M_5 \sim 10^{-10}$ GeV or less) in the interaction basis, will lead to the emergence of very small parameters in the first column of $M_D$ i.e., $ \widetilde{M_{D_{\alpha 1}}} \sim \left(\widetilde{M_{D_{ \alpha3}}} \frac{M_5}{M_6} \right) $ which is obtained after performing the diagonalization of the $\widetilde{M_R}$ matrix. Finally, due to this diagonalisation we write $\widetilde{M_R}$ as
\begin{equation}
                    \widetilde{M_{R}}=\begin{pmatrix}
                        \widetilde{M_1}  & 0  &  0 \\
                        0  & \widetilde{M_2}  & 0  \\
                        0 & 0  & \widetilde{M_3}   
                    \end{pmatrix}.
\end{equation}
This diagonalization however, make negligible change in (11) element in the above matrix in comparison to Eq. (28). Since, these Yukawa couplings ($\widetilde{Y_{\alpha 1}}$) of $\widetilde{N_1}$ (after diagonalization) are very small $\sim 10^{-14}$ or less, they do not disturb the neutrino oscillation observables obtained in the previous section.

For $\widetilde{N_1}$ to be suitable dark matter , a major requirement is that the particle should never had been in thermal equilibrium with the thermal bath particles. Then, the production proceeds non-thermally via feeble interactions with the bath particles. This is achieved by requiring very small couplings of RHN ($\widetilde{N_1}$) with the SM particles due to small $M_5$ and they will also lead to the production of the DM via decays and scatterings of the particles present in the thermal bath. Due to small couplings the population builds slowly and accumulates over time as the Universe expands and this results in the observed relic abundance.  However, they lead to very feeble interactions of $\widetilde{N_1}$ with the rest of the SM particles and also lead to its production in the early Universe, which is mainly dominated via 2-body decays of the gauge and higgs bosons. So, $\widetilde{N_1}$ yield, can be computed by solving 
the following Boltzmann equation \cite{Biswas:2016bfo}

\begin{align}
\frac{dY_{\widetilde{N_1}}}{dz} = \frac{2 M_{pl} z}{1.66 m_h^2} \frac{\sqrt{g_*(z)}}{g_s(z)}  \sum_{\alpha=e,\mu,\tau}  
\Big[ Y_Z^{eq} \big\langle\Gamma_{Z\to \widetilde{N_1} \nu_\alpha}\big\rangle  + Y_h^{eq} \big\langle\Gamma_{h\to \widetilde{N_1} \nu_\alpha}\big\rangle + Y_W^{eq} \big\langle\Gamma_{W^{\pm}\to \widetilde{N_1}\ell^{\pm}_{\alpha}}\big\rangle\Big],\label{1BE}
\end{align}
where $Y_{\widetilde{N_{1}}}(T)=n_{\widetilde{N_1}}(T)/s(T)$ and $z = m_h/T$. The above equation is solved under the condition that initially, the number density of $\widetilde{N_1}$ is zero to begin with i.e., $Y_{\widetilde{N_1}} (z \sim0)=0$, which is the standard assumption under the freeze-in scenario.  The quantity $\big\langle \Gamma_{A\to BC}\big\rangle$ represents the thermally averaged  decay width and is defined as:
\begin{eqnarray}
    \big\langle \Gamma_{A\to BC}\big\rangle = \frac{K_1(z)}{K_2(z)} \Gamma_{A\to BC}
\end{eqnarray}
where $K_1(z)$, and $K_2(z)$ are the modified Bessel functions of order $1$ and $2$, respectively. The function $g_*(z)$ is given by:

\begin{equation}
  \sqrt{g_*(z)} = \frac{g_s(z)}{\sqrt{g_\rho(z)}}\left(1 - \frac{1}{3} \frac{d \ln g_s(z)}{d \ln z}\right)  
\end{equation}
where $g_\rho(z)$ and $g_s(z)$ are the effective degrees of freedom related to energy density ($\rho$) and the entropy density (s) of the universe \cite{Biswas:2016bfo}, respectively. For the temperature regimes  Various decay rates that enter into the above Boltzmann equation, are given as: 
\begin{align}\label{GS1}
	\Gamma\left(h	\to \widetilde{N_{1}} \, \nu_{\alpha}  \right) & =  \frac{m_{h}\,\left| \widetilde{Y_{\alpha 1}} \right|^{2}}{32\,\pi}\left(1-\frac{\widetilde{M_{1}}^{2}}{m_{h}^{2}}\right)^{2}\ \nonumber,\\& \approx \frac{m_{h}\,\left| \widetilde{Y_{\alpha 1}} \right|^{2}}{32\,\pi},\nonumber \\
\end{align}
\begin{eqnarray} \label{wdecay}   
    \Gamma_{W^{\pm} \to \widetilde{N_1} \ell^{\pm}_\alpha} & =& \frac{1}{48\pi} m_W |\widetilde{Y_{\alpha 1}}|^2 f(\widetilde{M_{1}}^2 / m_W^2), \\
     \Gamma_{Z \to \bar{\widetilde{N_1}} \, \nu_\alpha\, +\, \widetilde{N_1}\, \bar{\nu}_\alpha} &=& \frac{1}{48\pi} m_Z |\widetilde{Y_{\alpha 1}}|^2 f(\widetilde{M_{1}}^2 / m_Z^2), \label{zdecay}   
\end{eqnarray}
where\quad
$$
f(a) = (1 - a)^2 (1 + 2/a).
$$
The approximations used in Eq.~(\ref{GS1}) are valid unless there is a near mass degeneracy between $\widetilde{N_1}$ and the Higgs field. It should be noted that the scatterings processes producing $\widetilde{N_1}$ are significantly suppressed ($\sim (\widetilde{Y_{\alpha 1}})^{4}$) and hence are excluded from the analysis. Back reactions involving $\widetilde{N_1}$ are also not included since initially the $\widetilde{N_1}$ number density is vanishingly small. For the same reason, terms proportional to $Y_{\widetilde{N_1}}$ are also dropped, which is a standard approximation for the freeze-in case~\cite{Hall:2009bx}. In order to compute the relic abundance ($\Omega_{\widetilde{N_{1}}} h^2$) of the sterile neutrino dark matter, one needs to find the value of its co-moving number density ($Y_{\widetilde{N_{1}}}$) at the present epoch ($T=T_{\infty}$ =2.73$K$). This value ($Y_{\widetilde{N_{1}}}(T=T_{\infty})$) is obtained by solving the Eq.\ \eqref{1BE} for the number density of $\widetilde{N_1}$. The expression of ($\Omega_{\widetilde{N_{1}}} h^2$) in terms of $Y_{\widetilde{N_{1}}}(T=T_{\infty})$ is given as~\cite{Edsjo:1997bg}:

\begin{figure}[h!]
  \centering
    \includegraphics[width=0.7\textwidth]{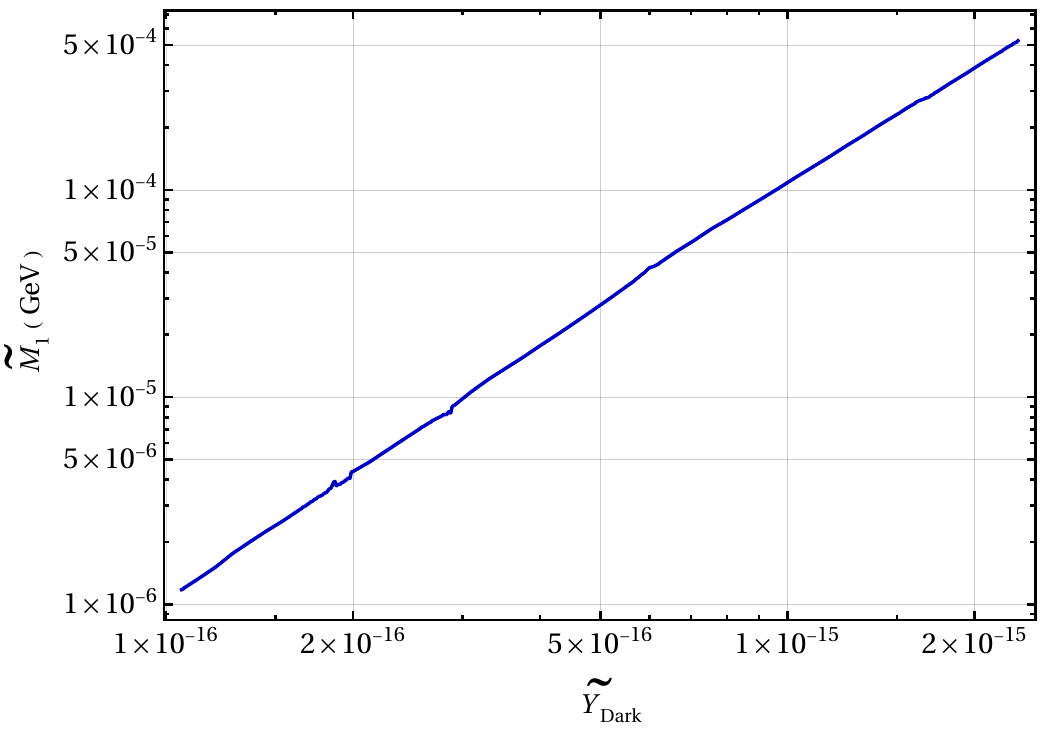}
    \caption{Variation of $\widetilde{M_{1}}$ versus $\widetilde{Y}_{Dark}$ that reproduces the observed relic density.}
    \label{fig:subfig1}
  
\end{figure}

\begin{equation}\label{relicbound}
\Omega_{\widetilde{N_1}}h^2 = 2.755\times 10^8 \bigg(\frac{\widetilde{M_{1}}}{\text{GeV}}\bigg)Y_{\widetilde{N_1}}(z_{\infty})
\end{equation}
where the allowed range for the value of the relic abundance is $0.118 \leq\Omega h^2 \leq 0.122$ as per Planck observations \cite{Planck2018Cosmo}. The freeze-in production of dark matter ($\widetilde{N_{1}}$) in this scenario has been studied quantitatively by numerically solving the Boltzmann Eq.~(\ref{1BE}). The unknown parameters in this analysis are the dark matter mass ($\widetilde{M_{1}}$) and its three Yukawa couplings ($\widetilde{Y_{i1}}$, where $i=1$ to $3$). However, in presence of soft breaking $M_5$ term, according to our previous discussions with coupling shown in Eq. (30), one may write $ \widetilde{Y_{ \alpha 1}} \approx \frac{M_5}{M_6} \widetilde{Y_{ \alpha 3}}$. After writing $\widetilde{Y}_{Dark} = \frac{M_5}{M_6}
\sqrt{\sum_\alpha|\widetilde{Y_{\alpha 3}|}^2}$ and using BP1
(consideration of BP1, is justified in the next section V) values in Table II for $\widetilde{Y_{\alpha 3}}$, effectively, there are only two unknown parameters: $\widetilde{M_1}$ and  $\widetilde{Y}_{Dark}$ due to $M_5$. It is desirable to study the relationship of these two free parameters with relic abundance requirements. We vary $\widetilde{M_{1}}$ and $\widetilde{Y}_{\text{Dark}}$ simultaneously and solve Eq.\ \eqref{1BE} to get $Y_{\widetilde{N_1}}(z_{\infty})$. Then, Eq.\ \eqref{relicbound} is used to calculate the relic abundance and only those pairs of $\widetilde{M_1}$, $\widetilde{Y}_{\text{Dark}}$ are kept that satisfy the Planck observations \cite{Planck:2018vyg} for the relic abundance. In Fig.\ \eqref{fig:subfig1}, we plot the respective values of these pair  as a blue line that provides the appropriate relic abundance. Using BP1 in Table \eqref{tab:bp_parameters}, $\widetilde{Y}_{Dark} = 5.72 \times 10^{-6} \left( \frac{M_5}{1 \, \text{GeV}} \right)$ corresponding to $\widetilde{M_2} \sim \widetilde{M_3} \sim 152 $ GeV. Then from Fig. -3, it follows that the possible allowed range of soft breaking term $M_5$ from dark matter relic abundance is given as : $1.78 \times 10^{-11} \; \mbox{GeV} \lesssim M_5 \lesssim 4.15 \times 10^{-10} \; \mbox{GeV}$ whereas the bound on dark matter mass $\widetilde{M_1}$ is : $ 10^{-6} \; \mbox{GeV} \lesssim \widetilde{M_1} \lesssim 5 \times 10^{-4} \; \mbox{GeV}$.

With higher allowed values of $M_5$ and $M_1$, considering Eq. (42), and $ \widetilde{M_{D_{\alpha 1}}} $ as discussed at the beginning of this section, from Eq. (32), the lightest active neutrino mass : $m_1 \approx 4.7\times 10^{-12}$ eV
is obtained corresponding to other parameters as in BP1, Table -II. This represents extremely hierarchical normal ordering scenario for light neutrino masses, satisfying required mass squared differences corresponding to neutrino oscillation data. The sum of all three neutrino masses is obtained as $\sum_i m_i \approx 0.058$ eV which is almost at the minimum possible value. This could be probed in future cosmological data. The current limits on the sum of light neutrino masses is $\sum_{i=1}^3 m_i <0.072$ eV at 95\% confidence level from analysis of BAO observations by DESI Collaboration \cite{DESI:2024hhd,DESI:2024mwx}. 

\section{Leptogenesis} \label{sec6}

In our setup, the exact degeneracy between the two heavy right handed neutrinos ($\widetilde{N_2}$ and $\widetilde{N_3}$) hints towards the fact that this framework could be utilized to provide an explanation of the observed baryonic asymmetry of the Universe via the Resonant Leptogenesis (RL) mechanism. This mechanism mainly utilizes the self-energy diagram \cite{Covi:1996wh,Pilaftsis:2003gt,Pilaftsis:2005rv, Pilaftsis:1997jf,Pilaftsis:1998pd} for the decays (as triangle diagram contribution is quite suppressed) and requires nearly degenerate heavy Majorana neutrinos which participate in the early $CP$ asymmetry creation. The leptonic asymmetry gets converted to the baryonic asymmetry via Sphaleron transitions. It turns out that it is possible to explain the baryonic asymmetry at a lower scale (below  TeV scale) for the heavy right handed neutrinos. Since, the present setup already accounts for the light neutrino observables of mass and mixing angles with only 3 complex Yuakwa couplings and one real parameter for the RHN masses at a low scale for the RHN masses, it is desirable to explore the resonant leptogenesis mechanism within this framework with a few parameters only.

It is rather interesting to observe from the benchmark points given in TABLE- \eqref{tab:bp_parameters} that the Yukawa couplings for the three flavors of the active neutrinos differ considerably with the $\nu_{\tau}$ couplings achieving the largest value among the three flavors. As we are considering the Leptogenesis scenario to be below the TeV scale so it is necessary to distinguish different flavors \cite{Nardi:2005hs} of the light neutrinos and their respective couplings with right handed neutrinos. So, in the analysis of leptogenesis, the effect of various flavors are taken into consideration individually in the leptonic asymmetry generation. The amount of flavored CP-asymmetry created in the early Universe is quantified by the $(\epsilon_{i \alpha})$ parameter \cite{Chauhan:2024jfq,Huang:2024azp,DeSimone:2007rw} which is defined as:
\begin{equation}
    \epsilon_{i \alpha}= \frac{\Gamma( \widetilde{N_i} \rightarrow L_\alpha \Phi^\dagger)-\Gamma( \widetilde{N_i} \rightarrow L_{\alpha}^C \Phi)}{\Gamma( \widetilde{N_i} \rightarrow L_\alpha \Phi^\dagger)+\Gamma( \widetilde{N_i} \rightarrow L_{\alpha}^C \Phi)} ;\quad \text{for $i=2,3$}
\end{equation}
where $\Gamma$'s are the decay widths of $\widetilde{N_i}$ and $\alpha$ is the flavor index. At the tree level the $CP$-asymmetry parameter $\epsilon_{i \alpha}$ vanishes. However, if one considers higher order contributions (say, at one-loop) then it is possible to obtain non-zero values of leptonic asymmetry (Eq.(50)) via the interference between the tree and self-energy diagrams in the $ \widetilde{N_i}$ decays. Once, a non-zero value of $\epsilon_{i \alpha}$ has been obtained, the $(B-L)$- conserving electroweak sphaleron transition processes convert this leptonic asymmetry to the baryonic asymmetry. Below the critical temperature ($T_c \sim 150$ GeV) the electroweak phase transition occurs and the sphaleron freeze-out takes place around $T\sim 131$ GeV \cite{Chauhan:2024jfq} which is slightly below $T_c$. 

In the RL mechanism, it is possible to obtain an enhanced value for $\epsilon_{i \alpha}$ if in the self-energy diagram, the intermediate state $\widetilde{ N_{j}}\, (j \neq i)$  is quasi-degenerate in mass with the initial state $ \widetilde{N_i}$. For the present scenario, this means that almost degenerate $\widetilde{N_{2}}$ and $\widetilde{N_{3}}$ are required, which is naturally facilitated by introducing soft-symmetry breaking parameter $M_3$ in the heavy Majorana neutrino mass matrix.

In order to obtain non-zero $CP$-asymmetry in Eq.(50), the imaginary part of the product of four couplings involved in the interference term of the tree level amplitude diagram and self energy diagram for the decay $\widetilde{N_i} \rightarrow L_\alpha \Phi^\dagger$, is required to be non-zero. However, with $Z_4$ symmetry shown in Table I for $M_D$ in Eq. (17) and corresponding $\widetilde{M_D}$ in Eqs. (29) and (30), $Y_{\alpha 2}=0$, for which imaginary part mentioned above, vanishes. However,  with some non-zero value of one of the $Y_{\alpha 2}$ as small soft $z_4$ symmetry breaking term,  non-zero imaginary part is obtained. Considering  $Y_{1 2}= k_1$ in $\widetilde{M_D}$ in Eqs. (29) where $k_1$ is some small real value - a few order smaller than  $|k|$ shown as the benchmark value in Table II, we rewrite $\widetilde{M_D}$ as
\begin{equation}
                    \widetilde{M_{D}}=v \widetilde{Y}  = v \begin{pmatrix}
                        0   & -\iota  \,   (k \sin{\theta} -k_1  \cos{\theta})  &  \,   (k \cos{\theta}+k_1  \sin{\theta})  \\
                        0  & -\iota  \, \alpha  \sin{\theta}   & \, \alpha \cos{\theta}  \\
                        0 & -\iota  \, \beta  \sin{\theta}   & \, \beta  \cos{\theta}    
                    \end{pmatrix}.
\end{equation}
in which $\widetilde{Y_{\alpha 1}}$ discussed in the previous section, has been ignored as those are too small to play any role in leptogenesis. For $\widetilde{Y}$ in Eq. (51), the imaginary part of the product of the couplings, is non-zero and for small $k_1 \sim 3 \times 10^{-7}$ say, which is about 3 order lesser than  $|k|$ as shown in Table II, there are very insignificant changes in neutrino masses and mixings and complex phase as shown in Table III and those remain within experimental $3 \sigma$ limits for $ \widetilde{Y}$ in Eq.(51).

In our work, it has been verified that only for heavy right handed neutrino fields around 152 GeV, it is possible to obtain the required baryonic asymmetry. However, this is around electro-weak scale. So we consider $(\widetilde{N_i} \rightarrow \nu_\alpha h^\dagger)$ instead of $(\widetilde{N_i} \rightarrow L_\alpha \Phi^\dagger)$ in our later discussions. Here, $\nu_\alpha$ denotes light neutrino with flavor $\alpha$ and $h$ is the SM higgs.

We consider the effective temperature dependent Higgs mass $m_h(T)$ due to interactions with hot plasma \cite{Giudice:2003jh,Elmfors:1993re,Klimov:1981ka,Comelli:1996vm,Cline:1993bd} in the early universe . This is given by 
\begin{equation}
{m_h(T)}^2= {m_h(0)}^2+ c\; T^2
\end{equation}
where  $c=\frac{3}{16} g^2+\frac{1}{2} g'^2+\frac{1}{4} y_t^2+\frac{1}{2}\lambda \sim 0.4 $ and $g=\frac{e}{\sin{\theta_w}}$ and $g'=\frac{e}{\cos{\theta_w}}$ are gauge couplings, $y_t$ is the top quark Yukawa coupling and $\lambda$ is the Higgs quartic coupling. The zero temperature Higgs mass $m_h(0)$ is zero above critical temperature of about 150 GeV. Thermal correction to heavy right handed neutrino mass and light neutrino mass are negligible due to small Yukawa couplings $\widetilde{Y}$ as considered in Eq. (51)  
and Table II. The two body decay width in the rest frame of decaying $\widetilde{N_i}$ is given by
\begin{equation}
\Gamma_{D_i}=\Gamma (\widetilde{N_i} \rightarrow \nu_\alpha h)= \frac{\widetilde{M_i} \left( \widetilde{ Y^\dagger} \widetilde{ Y} \right)_{ii}}{16 \pi}\left(1-m_h(T)^2/\widetilde{M_i}^2\right)^2\; .
\end{equation}

Considering the interference term of tree level and self energy one loop level amplitudes in Eq. (50), the $CP$ asymmetry parameter  \cite{Huang:2024azp,Chauhan:2024jfq,Chauhan:2021xus,Granelli:2020ysj,DeSimone:2007rw}  is  given by:
\begin{equation}
 \varepsilon_{i\alpha}
= \sum_{j\neq i}
\frac{\operatorname{Im}
\left[
(\widetilde{ Y}^\dagger)_{ i \alpha} (\widetilde{ Y})_{\alpha j}
\left( \widetilde{ Y^\dagger} \widetilde{ Y} \right)_{ij}
\right]}{ \left( \widetilde{ Y^\dagger} \widetilde{ Y} \right)_{ii} \left( \widetilde{ Y^\dagger} \widetilde{ Y} \right)_{jj}  }
\frac{
\widetilde{ M_i}  \; \Gamma_{D_j} (\widetilde{ M_i}^2 - \widetilde{ M_j}^2)
}{
(\widetilde{ M_i}^2 - \widetilde{ M_j}^2)^2 + (\widetilde{ M_i}\; \Gamma_{D_j})^2
}
\end{equation}
where the index $\alpha = e,\,\mu \, \tau$ and this accounts for $CP$ asymmetry due to different flavors of the active neutrinos in the final states in decay while the index $i=2 \, \text{and} \, 3$ denotes the RHNs and  $\widetilde{Y}$ is the Yukawa coupling matrix given by Eq. (51). From the form of the above equation it is evident that there exists a possibility to obtain a maximal enhancement ( resonance) if the condition 
$\widetilde{ M_{3}}- \widetilde{M_{2}}= \frac{\Gamma_{D_i}}{2}
$ is obeyed.
So, the differences between the masses of the heavy right handed neutrinos which participate in the asymmetry generation should be comparable to their decay widths.  In our case, such small difference is naturally provided by $M_3$ in Eq. (17) as discussed below Eq. (17) and one may consider  the soft-breaking term, $M_3$ as
\begin{equation}
M_3 \sim (\widetilde{ M_{3}}- \widetilde{M_{2}}) \sim \frac{\Gamma_{D_i}}{2} .
\end{equation}
One may note that as $\Gamma_{D_i}$ in Eq. (53) is temperature dependent, only near resonance condition could be obtained over a small range of the temperature $T$ for specific value of $M_3$ in the context of Eq.(55).

The required baryonic asymmetry could be obtained for the decay $\widetilde{N_i} \rightarrow \nu_\alpha h$ with right handed neutrino mass $\widetilde{ M_{3}} \sim \widetilde{M_{2}} = M_6 = 152 $ GeV with appropriate couplings as shown Eq. (51) and BPI in Table II. The wash out parameter 
$$ K_i = \frac{\Gamma_{D_i}}{H(T = \widetilde{M_{i}}) }$$
(where $T$ is the temperature of the universe and 
$H$ is the Hubble parameter defined as, $H(T)= \sqrt{\frac{4 \pi^3 g_{*}}{45}} \frac{T^2 }{M_{\text{Pl}}}$ where $M_{Pl}= 1.22 \times 10^{19}$ GeV is the Planck scale) is very large for the couplings in Table II and for $K_i >> 1$, such decay processes are in the strong wash out regime. Such strong washout is dominated by inverse decays and the washout due to various scattering processes are relatively much lesser. As we are considering leptogenesis around 152 GeV scale, different flavors of neutrinos are to be distinguished in considering wash out. 

With $z=M_{\widetilde{N_i}}/T$, the ratio of number density $n_{\widetilde{N_i}}(z)$  with respect to entropy density $s(z)$ and lepton asymmetry with respect to entropy density are defined as
\begin{equation}
Y_{\widetilde{N_i}}= \frac{n_{\widetilde{N_i}}(z)}{s(z)}\;\;, 
\;\;Y_{ \Delta L_\alpha}= \frac{n_{l_\alpha}(z)-n_{\bar{l_\alpha}}(z)}{s(z)}
\end{equation}
respectively where entropy density $s(z) = g^*\frac{2 \pi^2}{45} \frac{{M_{\widetilde{N_2}}}^3}{z^3}$ and effective number of degrees of freedom $g^* \sim 112$ as in SM after taking into account the heavy RHNs. The coupled Boltzman equations for $Y_{\widetilde{N_i}}$ and $Y_{ \Delta L_\alpha}$  are given by:   
\begin{align}
\frac{d Y_{\widetilde{N_i}}}{d z}
&=
-\frac{z}{H(\widetilde{M_2})\, s(z)}
\left[
\left(\frac{Y_{\widetilde{N_i}}}{Y_{\widetilde{N_i}}^{\mathrm{eq}}}-1\right)
\gamma_{D_i}
\right],
\\[1em]
\frac{d Y_{ \Delta L_\alpha}}{d z}
&=
\frac{z}{H(\widetilde{M_2})\, s(z)}
\Bigg[
\sum_{i}
\epsilon_{i\alpha}
\left(\frac{Y_{\widetilde{N_i}}}{Y_{\widetilde{N_i}}^{\mathrm{eq}}}-1\right)
\gamma_{D_i}
\nonumber
\\
&\hspace{2.8cm}
-
\frac{1}{2}
\frac{Y_{ \Delta L_\alpha}}{Y_\ell^{\mathrm{eq}}}
\sum_{i}
B_{i\alpha}\,
\gamma_{D_i}
\Bigg].
\end{align}
where the index $i=2,3$ is associated with the heavy right handed neutrinos and $\alpha=e,\mu,\tau$ keeps track of different flavors of the active neutrinos and the branching ratios 
$
B_{i\alpha}
=
\frac{|\widetilde{Y}_{\alpha i}|^2}
{(\widetilde{Y}^\dagger \widetilde{Y})_{ii}}
$. The thermally averaged decay widths $\gamma_{D_j} $ in above equations are defined as:
 \begin{eqnarray}
     \gamma_{D_j} := \gamma^{eq} ( \widetilde{N}_{j} \rightarrow h + \nu) + \gamma^{eq} (\widetilde{N}_{j} \rightarrow h + \bar{\nu}) =  {\Gamma_{D_j}}n_{N_j}^{eq} \frac{K_1(z)}{K_2 (z)}
 \end{eqnarray}
 where  $K_1$ and $K_2$ are the modified Bessel functions of first kind, $H(\widetilde{M_2})$ is the Hubble parameter evaluated at the mass-scale of the heavy right handed neutrino $\widetilde{N_2}$, ${\Gamma_{D_j}}$ is given by Eq. (53) and the ratio of the two Bessel functions acts as a time dilatation factor. Also, the equilibrium number density $(n^{eq}_{j})$ of  $\widetilde{N_j}$ is a function of temperature and could be written in terms of the variable $z$ as:

 \begin{equation}
     n^{eq}_{j} (z) = \frac{g_{j} \widetilde{M}_{j}^3}{2 \pi^2 z} K_{2}(z) 
 \end{equation}
where $g_{j}$ is the degree of freedom of the $\widetilde{N_j}$ with its mass $\widetilde{M_j}$. 

We discuss here, the numerical computations of the above Boltzman equations. Solving Eq.(57) and Eq.(58) numerically, one obtains three leptonic asymmetries $Y_{ \Delta L_\alpha}$ at different $z$ values. Following benchmark point $\textbf{BP1}$ as given in TABLE-\ref{tab:bp_parameters} which also satisfies the neutrino observables within $3 \sigma$ (see TABLE- \ref{tab:pmns_observables}), we have considered
mass $\widetilde{M_2} = 152$ GeV for the decaying right handed neutrino field $\widetilde{N_2}$. Following $\widetilde{Y}$ given in Eq. (51) with different parameter values given in  $\textbf{BP1}$ in Table II and soft breaking parameter $k_1 = 3 \times 10^{-7}$ as discussed after Eq. (51) and $M_3$ $\sim 0.4$ eV in Eq.(55) and using eq. (53)
for $\Gamma_{D_j}$, $\epsilon_{i\alpha}$ in Eq. (54) could be obtained which varies with temperature.
 We assume that there was no initial lepton number asymmetry and $Y_{ \Delta L_\alpha} = 0$  in the early Universe and all the particles were at some point, part of the thermal bath and Boltzmann equations are solved with the following initial conditions: $Y_{N_i} (z=z_{\text{in}}) = Y_{N_{i}}^{\text{eq}} (z=z_{\text{in}})$ and 
      $Y_{\Delta L_\alpha} (z=z_{\text{in}}) =0$, for $\alpha = e, \, \mu, \, \tau$.
\begin{figure}[h!]
  \centering
    \includegraphics[width=0.7\textwidth]{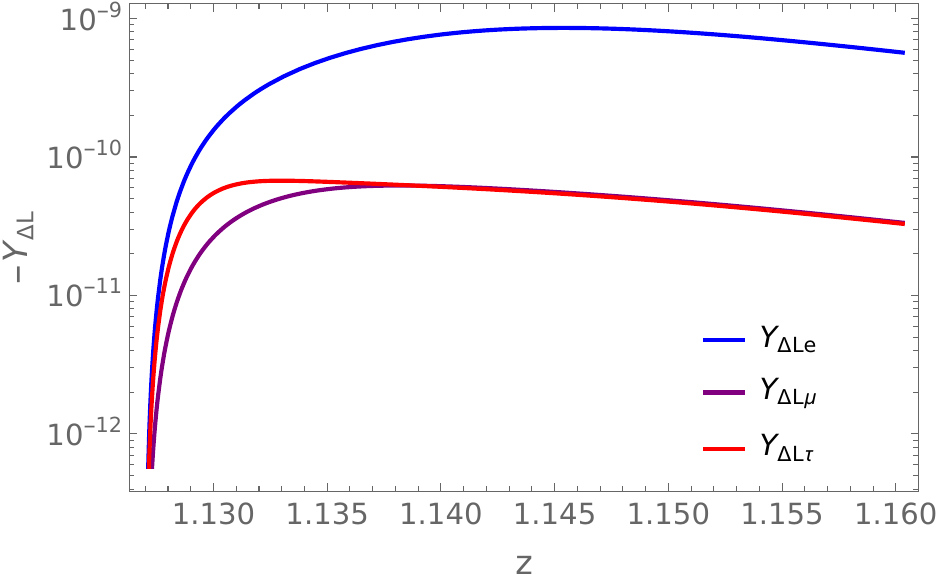}
    \caption{Variation of $-Y_{\Delta L }$ versus $z$ due to  $\widetilde{N_i}$ decays to  different flavors of light neutrinos.}
    \label{fig:subfig2}
  \end{figure}
The decay of $\widetilde{N_i}$ for $\widetilde{M_i} \sim 152$ GeV in Eq. (53) with $m_H(0)=125  $ GeV is highly phase space suppressed and the decay occurs for $T$  below 140 GeV and   
$z_{\text{in}} \sim 1.113 $ is kinematically possible. The effective conversion of leptonic asymmetry to baryonic asymmetry through sphaleron transition is possible up to the temperature of about 131 GeV and for that corresponding $z$ value is $z_{\text{final}} \sim 1.16 $. Solving Boltzman Eqs. (57) and (58) in the above range of $z$, the variation of $Y_{\Delta L \alpha} $ versus $z$ has been shown in Figure - 4 for three different flavor of light active neutrinos in the final states due to the decays of both $\widetilde{N_2}$ and $\widetilde{N_3}$. The total ratio of lepton asymmetry to entropy density is given by:
\begin{eqnarray} 
    Y_{\Delta L}= Y_{\Delta Le}+Y_{\Delta L \mu}+Y_{\Delta L \tau}
\end{eqnarray}
From Figure 4,  following values are found:
$Y_{\Delta Le} \approx -5.68 \times 10^{-10}$, $Y_{\Delta L \mu} \approx -3.29 \times 10^{-11}$ and  $Y_{\Delta L \tau} \approx  -3.29 \times 10^{-11}$. From those values, $Y_{\Delta L} \approx -6.34 \times 10^{-10}$.
In  \textbf{BP1}, since the Yukawa couplings for the electron neutrino are the smallest, so the inverse decay processes are subdominant for this particular case as the muon neutrino and tau neutrino couplings are larger, the inverse decay processes for them lead to considerable washout of the lepton asymmetry. So, $Y_{\Delta L e}$ gives the major contribution in $Y_{\Delta L}$. Initial rise of $Y_{\Delta L \tau}$ in comparison to $Y_{\Delta L \mu}$ in Fig 4 is due to the higher values of $\epsilon_{i\tau}$ in comparison to 
$\epsilon_{i\mu}$. 

Then, $Y_{\Delta L}$ gets converted to $Y_{ \Delta B}= \frac{n_{b}(z)-n_{\bar{b}}(z)}{s(z)}$ (which is baryonic number density asymmetry to entropy density ratio), via sphaleron transition and is given as:
\begin{equation}
Y_{\Delta B}= - \frac{28}{79} Y_{\Delta L} \approx 2.25 \times 10^{-10}\; .
\end{equation}
 It shows more matter than antimatter in the universe.
 One can easily obtain $Y_{\Delta  B} \approx 8.7 \times 10^{-11}$ which corresponds to the data from the Planck Collaboration \cite{Planck:2018vyg}, by slightly changing the values of either both or one of the soft symmetry breaking parameters like $k_1$ or $M_3$ -  either reducing slightly the soft symmetry breaking $k_1$ value from $3 \times 10^{-7}$ to about $10^{-7}$ or slightly increasing $M_3$ value to about 1 eV or changing both the parameters. 

 For getting matter domination over antimatter, $ Y_{\Delta B}$ is required to be positive and due to the relation in Eq. (62), $Y_{\Delta L}$ is required to be negative (as in Fig. 4)  and for that  $\epsilon_{i\alpha}$ in Eq.(54) is required to be negative. With $Z_4$ symmetry and with soft breaking $k_1$ , the form of Yukawa  couplings are shown in Eq. (51). The sign of $\epsilon_{i\alpha}$ depends on soft symmetry breaking term $k_1$. For real $k_1$ as considered in obtaining Fig. 4, the sign of $k_1$ is necessarily required to be positive, for negative $\epsilon_{i\alpha}$. If $k_1$ is considered to be complex, then a small negative imaginary part of the order of real part,  leads to a slightly more negative $\epsilon_{i\alpha}$ resulting in slightly increase in the magnitude of negative  value of $Y_{\Delta L}$ and for positive sign, there will be slight decrease in the magnitude of the negative  value of $Y_{\Delta L}$.
  
 One important point to note here that with effective thermal Higgs mass, the phase space for the decay of $\widetilde{N_i}$ is highly suppressed for $\widetilde{M_i}$ about 152 GeV, as is evident from the 
square term on the right hand side of Eq. (53). Because of this suppression, the value of $\Gamma_{D_i}$ is about two order lesser than that with the consideration of zero  Higgs mass and for this reason, the wash out due to inverse decay is lesser than what could have been with zero Higgs mass. This results in obtaining appropriate baryonic asymmetry. However, this suppression become much lesser for $\widetilde{M_i}$ somewhat higher than 152 GeV, for which, however, appropriate neutrino mass, mixing and $CP$ violating phase, can be obtained as shown in Table II and III. Because of lesser suppression of decay width in these cases, there will be more washout from inverse decays and it is very difficult to get required baryonic asymmetry.

\subsection{Searches for right handed neutrinos of masses around 152 GeV}
For  mass of right-handed neutrinos of 152 GeV, the heavy-light neutrino mixings are (following Eq.(8) and Eq. (30) and $\theta = \frac{\pi}{4}$ with BP1 in Table II):
\begin{eqnarray} 
|U_{\nu_e{\widetilde N_3}}|^2\approx \frac{|\widetilde{{M_D}_{13}}|^2}{|\widetilde{M_3}|^2}  \approx \frac{v^2 |k|^2}{ 2 M_6^2} \approx 6.6 \times 10^{-8} \nonumber \\
|U_{\nu_\mu{\widetilde N_3}}|^2\approx \frac{|\widetilde{{M_D}_{23}}|^2}{ |\widetilde{M_3}|^2}  \approx \frac{v^2 |\alpha|^2}{2 M_6^2} \approx 2.1 \times 10^{-7} \nonumber \\
|U_{\nu_\tau{\widetilde N_3}}|^2\approx \frac{|\widetilde{{M_D}_{33}}|^2}{|\widetilde{M_3}|^2}  \approx \frac{v^2 |\beta|^2}{2 M_6^2} \approx 7.1 \times 10^{-7}
\end{eqnarray}
If we replace in the above, $\widetilde{N_3}$ by $\widetilde{N_2}$ those are almost the same because of their quasi-degenerate masses. Replacing $\widetilde{N_3}$ by $\widetilde{N_1}$ in Eq. (63) leads to too small values because of soft symmetry breaking term $M_5$.
There are searches for $N_i$ decaying into $W^*$ (off-shell) boson  and charged lepton $\it{l}$, in the trilepton signal process: 
$pp\rightarrow W^* \rightarrow \it{l} \widetilde{N_i} \rightarrow \it{l}(\it{l} W^*) \rightarrow \it{l}(\it{l}\it{l} \nu)$
using the CMS detector at LHC \cite{CMS:2018jxx}. The experimental upper bound for $\widetilde{M_i} \sim 152$ GeV, as shown in Fig . 2 of \cite{CMS:2018jxx} is   $|U_{\nu_e{\widetilde N_3}}|^2 \lesssim 10^{-2}$ and $|U_{\nu_\mu{\widetilde N_3}}|^2 \lesssim 10^{-2}$. But, the corresponding values obtained in this work, which are shown in Eq. (63), are much smaller than the experimental bound. So, $\widetilde{N_2}$ and $\widetilde{N_3}$ considered here, could exist. However, in future electron proton colliders \cite{Antusch:2016ejd} in which SM QCD backgrounds are lesser than at $pp$ colliders at LHC, there is scope to improve the upper bounds in the lepton number violating signal process: $pe^- \rightarrow \mu^+ jjj$ with multijets in the final states.  As shown in Fig. 5 of  \cite{Antusch:2016ejd}, the upper limits on these mixings could improve significantly and could be lowered to about $2 \times 10^{-6}$ for the first two mixings in Eq. (63). There could be  some variations of the couplings (corresponding to neutrino mass square differences, mixings, $CP$ phase values at higher confidence level) from  those shown in BP1 in Table II for right handed neutrino mass at about 152 GeV and  in future, in the electron proton colliders, it could be possible to observe  $\widetilde{N_2}$ and $\widetilde{N_3}$ discussed here. However, as the centre of mass energy is higher in proton colliders at LHC, the high luminosity HL-LHC could play some complementary role along with $e p$ colliders in searching heavy right handed neutrinos around 152 GeV.  

\section{Conclusion} \label{sec7}

In the Type-1 seesaw mechanism, with $Z_4$ symmetry, it is possible to accommodate : (1) the neutrino oscillation data at $3\sigma$ confidence level, (2) dark matter requirements via freeze-in mechanism and (3) observed baryonic asymmetry through resonant leptogenesis near electroweak scale. Furthermore, it is interesting to note that for (1) only three complex parameters ($k, \, \alpha$ and $\beta$) in $\widetilde{M_D}$ and one real mass parameter ($M_6$) in $\widetilde{M_R}$ is required, for (2) one needs further a real soft-symmetry breaking parameter ($M_5$) and also for (3) one requires further two real soft-symmetry ($k_1$ and $M_3$) breaking parameters.

In general, for the massless texture in Type-1 seesaw mechanism, some constraint conditions on the parameters (resulting in fine tuning) of the seesaw mass matrix, are required. However, we have discussed where no such constraint conditions are required. In that context, we showed that $Z_4$ symmetry could be considered which further eliminates a few terms from the massless texture. 

To get massive light neutrinos one has to consider one-loop corrections to the seesaw-mass matrix. Although, in general, one-loop corrections to  the $\widetilde{M_{L}}$ block is dominant \cite{Grimus:2002nk,AristizabalSierra:2011mn} but with two quasi-degenerate heavy RHNs, one-loop corrections to the $\widetilde{M_D}$ block (shown in Appendix (\ref{App})) becomes dominant and this leads to three light neutrinos with appropriate masses, mixing and $CP$ violating phase as shown in Table II and III. 

  Soft $Z_4$ symmetry breaking breaking parameters are naturally small and play important role while considering  dark matter and resonant leptogenesis. One such parameter $M_5$ leads to very small non-zero couplings for $\widetilde{N_1}$ and  it is possible to consider $\widetilde{N_1}$  as a dark matter candidate via freeze-in mechanism.  Similarly,  parameter $k_1$ is useful in getting non-zero $CP$ asymmetry and small $M_3$   break the mass degeneracy between $\widetilde{N_2}$ and $\widetilde{N_3}$ and due to its smallness, the mechanism of Resonant Leptogenesis could be invoked naturally to address the baryonic asymmetry problem. Also, the sign of real $k_1$, as discussed in the previous section, is necessarily positive for the domination of matter over antimatter.

From Table II, one can see that two heavy right handed neutrino masses  related to $M_6$,  is not fixed by the requirement of satisfying neutrtino oscillation data. However, when leptogenesis is considered through the  decays of heavy two right handed neutrinos and thermal Higgs mass is taken into account, the two heavy right-handed neutrino masses (related to $M_6$) are required to be around 152 GeV. 
However,   heavy-light neutrino mixings shown in Eq. (63) in this work, are small for 152 GeV RHN, for present detection. But  there is scope of detecting such heavy neutrinos in future, possibly in the electron proton colliders \cite{Antusch:2016ejd}. 

\hspace*{\fill}

\textbf{Acknowledgement}: Kunal Pandey would like to thank Imtiyaz Ahmad Bhat for useful discussions and fruitful suggestions.

\appendix

\section{Expressions for the various one-loop corrections:} \label{App}

These are the various one-loop corrections whose analytical forms are given below:

\begin{align}
\epsilon_{i2}^{Z^{(1)}} &= \frac{1}{64 \, \cos^2\theta_W \, \pi^2 \, \widetilde{M_{D_{i2}}}^2 \left( \widetilde{M_{D_{i2}}} - \widetilde{M_{N_2}} \right) \widetilde{M_{N_2}^2}} \, g^2 \left( \widetilde{M_{D_{12}}}^2 + \widetilde{M_{D_{22}}}^2 + \widetilde{M_{D_{32}}}^2 \right) \\ \nonumber
&\times \Bigg[ \widetilde{M_{D_{i2}}}^2 \left( -2 M_Z^2 + 3 \widetilde{M_{D_{i2}}}^2 + 6 \widetilde{M_{D_{i2}}} \widetilde{M_{N_2}} + 2 \widetilde{M_{N_2}}^2 \right) \\ \nonumber
&\quad + 2 \widetilde{M_{D_{i2}}}^3 \left( \widetilde{M_{D_{i2}}} + 2 \widetilde{M_{N_2}} \right) \log \left( \frac{\mu^2}{\widetilde{M_{N_2}}^2} \right) \\ \nonumber
&\quad + \left( 2 M_Z^2 \widetilde{M_{D_{i2}}}^2 + \widetilde{M_{D_{i2}}}^4 + 2 \widetilde{M_{D_{i2}}}^3 \widetilde{M_{N_2}} - \left( -M_Z^2 + \widetilde{M_{N_2}}^2 \right)^2 - 2 \widetilde{M_{D_{i2}}} \left( -M_Z^2 \widetilde{M_{N_2}} + \widetilde{M_{N_2}}^3 \right) \right) \log \left( \frac{\widetilde{M_{N_2}}^2}{M_Z^2} \right) \\ \nonumber
&\quad + 2 \left( -M_Z^2 + \widetilde{M_{D_{i2}}}^2 + 2 \widetilde{M_{D_{i2}}} \widetilde{M_{N_2}} + \widetilde{M_{N_2}}^2 \right) f_{i2} \log \left( \frac{ M_Z^2 - \widetilde{M_{D_{i2}}}^2 + \widetilde{M_{N_2}}^2 + f_{i2} }{ 2 M_Z \widetilde{M_{N_2}} } \right) \Bigg],
\end{align}

\begin{align}
f_{i2} &= \sqrt{ \widetilde{M_{D_{i2}}}^4 + \left( -M_Z^2 + \widetilde{M_{N_2}}^2 \right)^2 - 2 \widetilde{M_{D_{i2}}}^2 \left( M_Z^2 + \widetilde{M_{N_2}}^2 \right) }.
\end{align}

\begin{align}
\epsilon_{i2}^{Z^{(2)}} &= \frac{g^2 \, \widetilde{M_{D_{i3}}} \left( \widetilde{M_{D_{13}}} \widetilde{M_{D_{12}}}^\ast + \widetilde{M_{D_{23}}} \widetilde{M_{D_{22}}}^\ast + \widetilde{M_{D_{33}}} \widetilde{M_{D_{32}}}^\ast \right) \left( \widetilde{M_{D_{i2}}} + \widetilde{M_{N_2}} \right)}{64 \, \cos^2\theta_W \, \pi^2 \, \widetilde{M_{D_{i2}}}^3 \, \widetilde{M_{N_3}}^2 \left( \widetilde{M_{D_{i2}}}^2 - \widetilde{M_{N_3}}^2 \right)} \\ \nonumber
&\times \Bigg[ \widetilde{M_{D_{i2}}}^2 \left( -2 M_Z^2 + 3 \widetilde{M_{D_{i2}}}^2 + 2 \widetilde{M_{N_2}}^2 + 6 \widetilde{M_{D_{i2}}} \widetilde{M_{N_3}} \right) \\ \nonumber
&\quad + 2 \widetilde{M_{D_{i2}}}^3 \left( \widetilde{M_{D_{i2}}} + 2 \widetilde{M_{N_3}} \right) \log \left( \frac{\mu^2}{\widetilde{M_{N_2}}^2} \right) \\ \nonumber
&\quad + \left( 2 M_Z^2 \widetilde{M_{D_{i2}}}^2 + \widetilde{M_{D_{i2}}}^4 - \left( -M_Z^2 + \widetilde{M_{N_2}}^2 \right)^2 + 2 \widetilde{M_{D_{i2}}}^3 \widetilde{M_{N_3}} + 2 \widetilde{M_{D_{i2}}} \left( M_Z^2 - \widetilde{M_{N_2}}^2 \right) \widetilde{M_{N_3}} \right) \log \left( \frac{\widetilde{M_{N_2}}^2}{M_Z^2} \right) \\ \nonumber
&\quad + 2 f_{i2} \left( -M_Z^2 + \widetilde{M_{D_{i2}}}^2 + \widetilde{M_{N_2}}^2 + 2 \widetilde{M_{D_{i2}}} \widetilde{M_{N_3}} \right) \log \left( \frac{ M_Z^2 - \widetilde{M_{D_{i2}}}^2 + \widetilde{M_{N_2}}^2 + f_{i2} }{ 2 M_Z \widetilde{M_{N_2}} } \right) \Bigg],
\end{align}

\begin{align}
f_{i2} &= \sqrt{ \widetilde{M_{D_{i2}}}^4 + \left( -M_Z^2 + \widetilde{M_{N_2}}^2 \right)^2 - 2 \widetilde{M_{D_{i2}}}^2 \left( M_Z^2 + \widetilde{M_{N_2}}^2 \right) }.
\end{align}

\begin{align}
\epsilon_{i2}^{H^{(1)}} &= \frac{v \left( |\widetilde{Y_{12}}|^2 + |\widetilde{Y_{22}}|^2 + |\widetilde{Y_{32}}|^2 \right) \widetilde{Y_{i2}}}{64 \pi^2 \widetilde{M_{D_{i2}}}^3 \widetilde{M_{N_2}}} \\ \nonumber
&\times \Bigg[ 2 \widetilde{M_{D_{i2}}}^2 \widetilde{M_{N_2}} \left( 4 \widetilde{M_{D_{i2}}} + \widetilde{M_{N_2}} \right) 
- 2 \left( -M_h^2 + \widetilde{M_{D_{i2}}}^2 \right)^2 \log \left( \frac{M_h^2}{M_h^2 - \widetilde{M_{D_{i2}}}^2} \right) \\ \nonumber
&\quad + \left( -2 M_h^2 \widetilde{M_{D_{i2}}}^2 + \widetilde{M_{D_{i2}}}^4 - 2 \widetilde{M_{D_{i2}}}^3 \widetilde{M_{N_2}} + \left( M_h^2 - \widetilde{M_{N_2}}^2 \right)^2 + \widetilde{M_{D_{i2}}} \left( -2 M_h^2 \widetilde{M_{N_2}} + 2 \widetilde{M_{N_2}}^3 \right) \right) \log \left( \frac{M_h^2}{\widetilde{M_{N_2}}^2} \right) \\ \nonumber
&\quad + 4 \widetilde{M_{D_{i2}}}^3 \widetilde{M_{N_2}} \log \left( \frac{\mu^2}{\widetilde{M_{N_2}}^2} \right) \\ \nonumber
&\quad + 2 f_{i2} \left( -M_h^2 + \widetilde{M_{D_{i2}}}^2 + 2 \widetilde{M_{D_{i2}}} \widetilde{M_{N_2}} + \widetilde{M_{N_2}}^2 \right) 
\log \left( \frac{ M_h^2 - \widetilde{M_{D_{i2}}}^2 + \widetilde{M_{N_2}}^2 + f_{i2} }{ 2 M_h \widetilde{M_{N_2}} } \right) \Bigg],
\end{align}

\begin{align}
f_{i2} &= \sqrt{ \widetilde{M_{D_{i2}}}^4 + \left( M_h^2 - \widetilde{M_{N_2}}^2 \right)^2 - 2 \widetilde{M_{D_{i2}}}^2 \left( M_h^2 + \widetilde{M_{N_2}}^2 \right) }.
\end{align}

\begin{align}
\epsilon_{i2}^{H^{(2)}} &= \frac{v \, \widetilde{Y_{i3}} \left( \widetilde{Y_{12}} \widetilde{Y_{13}}^\ast + \widetilde{Y_{22}} \widetilde{Y_{23}}^\ast + \widetilde{Y_{32}} \widetilde{Y_{33}}^\ast \right)}{64 \pi^2 \widetilde{M_{D_{i2}}}^3 \widetilde{M_{N_3}}} \\ \nonumber
&\times \Bigg[ 2 \widetilde{M_{D_{i2}}}^2 \widetilde{M_{N_3}} \left( 4 \widetilde{M_{D_{i2}}} + \widetilde{M_{N_3}} \right) 
- 2 \left( M_h^2 - \widetilde{M_{D_{i2}}}^2 \right)^2 \log \left( \frac{M_h^2}{M_h^2 - \widetilde{M_{D_{i2}}}^2} \right) \\ \nonumber
&\quad + \left( -2 M_h^2 \widetilde{M_{D_{i2}}}^2 + \widetilde{M_{D_{i2}}}^4 - 2 \widetilde{M_{D_{i2}}}^3 \widetilde{M_{N_3}} + \left( M_h^2 - \widetilde{M_{N_3}}^2 \right)^2 + \widetilde{M_{D_{i2}}} \left( -2 M_h^2 \widetilde{M_{N_3}} + 2 \widetilde{M_{N_3}}^3 \right) \right) \log \left( \frac{M_h^2}{\widetilde{M_{N_3}}^2} \right) \\ \nonumber
&\quad + 4 \widetilde{M_{D_{i2}}}^3 \widetilde{M_{N_3}} \log \left( \frac{\mu^2}{\widetilde{M_{N_3}}^2} \right) \\ \nonumber
&\quad + 2 f_{i3} \left( -M_h^2 + \widetilde{M_{D_{i2}}}^2 + 2 \widetilde{M_{D_{i2}}} \widetilde{M_{N_3}} + \widetilde{M_{N_3}}^2 \right) 
\log \left( \frac{ M_h^2 - \widetilde{M_{D_{i2}}}^2 + \widetilde{M_{N_3}}^2 + f_{i3} }{ 2 M_h \widetilde{M_{N_3}} } \right) \Bigg],
\end{align}

\begin{align}
f_{i3} &= \sqrt{ \widetilde{M_{D_{i2}}}^4 + \left( M_h^2 - \widetilde{M_{N_3}}^2 \right)^2 - 2 \widetilde{M_{D_{i2}}}^2 \left( M_h^2 + \widetilde{M_{N_3}}^2 \right) }.
\end{align}

\begin{align}
\epsilon_{i2}^W &= \frac{g^2}{32 \pi^2 \widetilde{M_{D_{i2}}}^2 \left( \widetilde{M_{D_{i2}}} - \widetilde{M_{N_2}} \right)} \\ \nonumber
&\times \Bigg[ \widetilde{M_{D_{i2}}}^2 \left( -2 M_W^2 + 3 \widetilde{M_{D_{i2}}}^2 + 2 m_{\ell_i}^2 \right) 
+ 2 \widetilde{M_{D_{i2}}}^4 \log \left( \frac{\mu^2}{m_{\ell_i}^2} \right) \\ \nonumber
&\quad + \left( 2 M_W^2 \widetilde{M_{D_{i2}}}^2 + \widetilde{M_{D_{i2}}}^4 - \left( -M_W^2 + m_{\ell_i}^2 \right)^2 \right) \log \left( \frac{m_{\ell_i}^2}{M_W^2} \right) \\ \nonumber
&\quad + 2 f_i^W \left( -M_W^2 + \widetilde{M_{D_{i2}}}^2 + m_{\ell_i}^2 \right) \log \left( \frac{ M_W^2 - \widetilde{M_{D_{i2}}}^2 + m_{\ell_i}^2 + f_i^W }{ 2 M_W m_{\ell_i} } \right) \Bigg],
\end{align}

\begin{align}
f_i^W &= \sqrt{ \widetilde{M_{D_{i2}}}^4 + \left( -M_W^2 + m_{\ell_i}^2 \right)^2 - 2 \widetilde{M_{D_{i2}}}^2 \left( M_W^2 + m_{\ell_i}^2 \right) }.
\end{align}

\begin{align}
\epsilon_{i3}^{Z^{(1)}} &= \frac{g^2 \left( \widetilde{M_{D_{13}}}^2 + \widetilde{M_{D_{23}}}^2 + \widetilde{M_{D_{33}}}^2 \right)}{64 \, \cos^2\theta_W \, \pi^2 \, \widetilde{M_{D_{i3}}}^2 \left( \widetilde{M_{D_{i3}}} - \widetilde{M_{N_3}} \right) \widetilde{M_{N_3}}^2} \\ \nonumber
&\times \Bigg[ \widetilde{M_{D_{i3}}}^2 \left( -2 M_Z^2 + 3 \widetilde{M_{D_{i3}}}^2 + 6 \widetilde{M_{D_{i3}}} \widetilde{M_{N_3}} + 2 \widetilde{M_{N_3}}^2 \right) \\ \nonumber
&\quad + 2 \widetilde{M_{D_{i3}}}^3 \left( \widetilde{M_{D_{i3}}} + 2 \widetilde{M_{N_3}} \right) \log \left( \frac{\mu^2}{\widetilde{M_{N_3}}^2} \right) \\ \nonumber
&\quad + \left( 2 M_Z^2 \widetilde{M_{D_{i3}}}^2 + \widetilde{M_{D_{i3}}}^4 + 2 \widetilde{M_{D_{i3}}}^3 \widetilde{M_{N_3}} - \left( -M_Z^2 + \widetilde{M_{N_3}}^2 \right)^2 - 2 \widetilde{M_{D_{i3}}} \left( -M_Z^2 \widetilde{M_{N_3}} + \widetilde{M_{N_3}}^3 \right) \right) \log \left( \frac{\widetilde{M_{N_3}}^2}{M_Z^2} \right) \\ \nonumber
&\quad + 2 f_{i3} \left( -M_Z^2 + \widetilde{M_{D_{i3}}}^2 + 2 \widetilde{M_{D_{i3}}} \widetilde{M_{N_3}} + \widetilde{M_{N_3}}^2 \right) \log \left( \frac{ M_Z^2 - \widetilde{M_{D_{i3}}}^2 + \widetilde{M_{N_3}}^2 + f_{i3} }{ 2 M_Z \widetilde{M_{N_3}} } \right) \Bigg],
\end{align}

\begin{align}
f_{i3} &= \sqrt{ \widetilde{M_{D_{i3}}}^4 + \left( -M_Z^2 + \widetilde{M_{N_3}}^2 \right)^2 - 2 \widetilde{M_{D_{i3}}}^2 \left( M_Z^2 + \widetilde{M_{N_3}}^2 \right) }.
\end{align}

\begin{align}
\epsilon_{i3}^{Z^{(2)}} &= \frac{\left( \widetilde{M_{D_{12}}} \widetilde{M_{D_{13}}} + \widetilde{M_{D_{22}}} \widetilde{M_{D_{23}}} + \widetilde{M_{D_{32}}} \widetilde{M_{D_{33}}} \right) \widetilde{M_{D_{i2}}}} {16 \pi^2 \widetilde{M_{D_{i3}}}^3 \widetilde{M_{N_2}}^2 \left( \widetilde{M_{D_{i3}}} - \widetilde{M_{N_3}} \right)} \\ \nonumber
&\times \Bigg[ \widetilde{M_{D_{i3}}}^2 \left( -2 M_Z^2 + 3 \widetilde{M_{D_{i3}}}^2 + 6 \widetilde{M_{D_{i3}}} \widetilde{M_{N_2}} + 2 \widetilde{M_{N_2}}^2 \right) \\ \nonumber
&\quad + 2 \widetilde{M_{D_{i3}}}^3 \left( \widetilde{M_{D_{i3}}} + 2 \widetilde{M_{N_2}} \right) \log \left( \frac{\mu^2}{\widetilde{M_{N_2}}^2} \right) \\ \nonumber
&\quad + \left( 2 M_Z^2 \widetilde{M_{D_{i3}}}^2 + \widetilde{M_{D_{i3}}}^4 + 2 \widetilde{M_{D_{i3}}}^3 \widetilde{M_{N_2}} + 2 \widetilde{M_{D_{i3}}} \widetilde{M_{N_2}} \left( M_Z^2 - \widetilde{M_{N_2}}^2 \right) - \left( M_Z^2 - \widetilde{M_{N_2}}^2 \right)^2 \right) \log \left( \frac{\widetilde{M_{N_2}}^2}{M_Z^2} \right) \\ \nonumber
&\quad + 2 f_{i32} \left( -M_Z^2 + \widetilde{M_{D_{i3}}}^2 + 2 \widetilde{M_{D_{i3}}} \widetilde{M_{N_2}} + \widetilde{M_{N_2}}^2 \right) 
\log \left( \frac{ M_Z^2 - \widetilde{M_{D_{i3}}}^2 + \widetilde{M_{N_2}}^2 + f_{i32} }{ 2 M_Z \widetilde{M_{N_2}} } \right) \Bigg],
\end{align}

\begin{align}
f_{i32} &= \sqrt{ \widetilde{M_{D_{i3}}}^4 + \left( M_Z^2 - \widetilde{M_{N_2}}^2 \right)^2 - 2 \widetilde{M_{D_{i3}}}^2 \left( M_Z^2 + \widetilde{M_{N_2}}^2 \right) }.
\end{align}

\begin{align}
\epsilon_{i3}^{H^{(1)}} &= \frac{v \, \widetilde{Y_{i3}} \left( |\widetilde{Y_{13}}|^2 + |\widetilde{Y_{23}}|^2 + |\widetilde{Y_{33}}|^2 \right)}{64 \pi^2 \widetilde{M_{D_{i3}}}^3 \widetilde{M_{N_3}}} \\ \nonumber
&\times \Bigg[ 2 \widetilde{M_{D_{i3}}}^2 \widetilde{M_{N_3}} \left( 4 \widetilde{M_{D_{i3}}} + \widetilde{M_{N_3}} \right) 
- 2 \left( M_h^2 - \widetilde{M_{D_{i3}}}^2 \right)^2 \log \left( \frac{M_h^2}{M_h^2 - \widetilde{M_{D_{i3}}}^2} \right) \\ \nonumber
&\quad + \left( -2 M_h^2 \widetilde{M_{D_{i3}}}^2 + \widetilde{M_{D_{i3}}}^4 - 2 \widetilde{M_{D_{i3}}}^3 \widetilde{M_{N_3}} + \left( M_h^2 - \widetilde{M_{N_3}}^2 \right)^2 + \widetilde{M_{D_{i3}}} \left( -2 M_h^2 \widetilde{M_{N_3}} + 2 \widetilde{M_{N_3}}^3 \right) \right) \log \left( \frac{M_h^2}{\widetilde{M_{N_3}}^2} \right) \\ \nonumber
&\quad + 4 \widetilde{M_{D_{i3}}}^3 \widetilde{M_{N_3}} \log \left( \frac{\mu^2}{\widetilde{M_{N_3}}^2} \right) \\ \nonumber
&\quad + 2 f_{i31} \left( -M_h^2 + \widetilde{M_{D_{i3}}}^2 + 2 \widetilde{M_{D_{i3}}} \widetilde{M_{N_3}} + \widetilde{M_{N_3}}^2 \right) 
\log \left( \frac{ M_h^2 - \widetilde{M_{D_{i3}}}^2 + \widetilde{M_{N_3}}^2 + f_{i31} }{ 2 M_h \widetilde{M_{N_3}} } \right) \Bigg],
\end{align}

\begin{align}
f_{i31} &= \sqrt{ \widetilde{M_{D_{i3}}}^4 + \left( M_h^2 - \widetilde{M_{N_3}}^2 \right)^2 - 2 \widetilde{M_{D_{i3}}}^2 \left( M_h^2 + \widetilde{M_{N_3}}^2 \right) }.
\end{align}

\begin{align}
\epsilon_{i3}^{H^{(2)}} &= \frac{v \, \widetilde{Y_{i2}} \left( \widetilde{Y_{12}}^\ast \widetilde{Y_{13}} + \widetilde{Y_{22}}^\ast \widetilde{Y_{23}} + \widetilde{Y_{32}}^\ast \widetilde{Y_{33}} \right)}{64 \pi^2 \widetilde{M_{D_{i3}}}^3 \widetilde{M_{N_2}}} \\ \nonumber
&\times \Bigg[ 2 \widetilde{M_{D_{i3}}}^2 \widetilde{M_{N_2}} \left( 4 \widetilde{M_{D_{i3}}} + \widetilde{M_{N_2}} \right) 
- 2 \left( M_h^2 - \widetilde{M_{D_{i3}}}^2 \right)^2 \log \left( \frac{M_h^2}{M_h^2 - \widetilde{M_{D_{i3}}}^2} \right) \\ \nonumber
&\quad + \left( -2 M_h^2 \widetilde{M_{D_{i3}}}^2 + \widetilde{M_{D_{i3}}}^4 - 2 \widetilde{M_{D_{i3}}}^3 \widetilde{M_{N_2}} + \left( M_h^2 - \widetilde{M_{N_2}}^2 \right)^2 + \widetilde{M_{D_{i3}}} \left( -2 M_h^2 \widetilde{M_{N_2}} + 2 \widetilde{M_{N_2}}^3 \right) \right) \log \left( \frac{M_h^2}{\widetilde{M_{N_2}}^2} \right) \\ \nonumber
&\quad + 4 \widetilde{M_{D_{i3}}}^3 \widetilde{M_{N_2}} \log \left( \frac{\mu^2}{\widetilde{M_{N_2}}^2} \right) \\ \nonumber
&\quad + 2 f_{i32} \left( -M_h^2 + \widetilde{M_{D_{i3}}}^2 + 2 \widetilde{M_{D_{i3}}} \widetilde{M_{N_2}} + \widetilde{M_{N_2}}^2 \right) 
\log \left( \frac{ M_h^2 - \widetilde{M_{D_{i3}}}^2 + \widetilde{M_{N_2}}^2 + f_{i32} }{ 2 M_h \widetilde{M_{N_2}} } \right) \Bigg],
\end{align}

\begin{align}
f_{i32} &= \sqrt{ \widetilde{M_{D_{i3}}}^4 + \left( M_h^2 - \widetilde{M_{N_2}}^2 \right)^2 - 2 \widetilde{M_{D_{i3}}}^2 \left( M_h^2 + \widetilde{M_{N_2}}^2 \right) }.
\end{align}

\begin{align}
\epsilon_{i3}^W &= \frac{g^2}{32 \pi^2 \widetilde{M_{D_{i3}}}^2 \left( \widetilde{M_{D_{i3}}} - \widetilde{M_{N_3}} \right)} \\ \nonumber
&\times \Bigg[ \widetilde{M_{D_{i3}}}^2 \left( -2 M_W^2 + 3 \widetilde{M_{D_{i3}}}^2 + 2 m_{\ell_i}^2 \right) 
+ 2 \widetilde{M_{D_{i3}}}^4 \log \left( \frac{\mu^2}{m_{\ell_i}^2} \right) \\ \nonumber
&\quad + \left( 2 M_W^2 \widetilde{M_{D_{i3}}}^2 + \widetilde{M_{D_{i3}}}^4 - \left( -M_W^2 + m_{\ell_i}^2 \right)^2 \right) 
\log \left( \frac{m_{\ell_i}^2}{M_W^2} \right) \\ \nonumber
&\quad + 2 f_{i3} \left( -M_W^2 + \widetilde{M_{D_{i3}}}^2 + m_{\ell_i}^2 \right) 
\log \left( \frac{ M_W^2 - \widetilde{M_{D_{i3}}}^2 + m_{\ell_i}^2 + f_{i3} }{ 2 M_W m_{\ell_i} } \right) \Bigg],
\end{align}

\begin{align}
f_{i3} &= \sqrt{ \widetilde{M_{D_{i3}}}^4 + \left( -M_W^2 + m_{\ell_i}^2 \right)^2 - 2 \widetilde{M_{D_{i3}}}^2 \left( M_W^2 + m_{\ell_i}^2 \right) }.
\end{align}

\begin{align}
\epsilon_{i 3}^{\phi^3 (2)} &= \frac{v}{8} \left( 
\widetilde{Y_{12}}^* \widetilde{Y_{13}} + \widetilde{Y_{22}}^* \widetilde{Y_{23}} + \widetilde{Y_{32}}^* \widetilde{Y_{33}} 
\right) \widetilde{Y_{i2}} \\
&\times \Bigg[
\frac{\widetilde{M_{D_{i3}}}^2 \left( M_Z^4 - \widetilde{M_{N_2}}^4 - 2 M_Z^2 \widetilde{M_{N_2}}^2 \log \frac{M_Z^2}{\widetilde{M_{N_2}}^2} \right)}
{32 \pi^2 \left( M_Z^2 - \widetilde{M_{N_2}}^2 \right)^3} \nonumber \\
&\quad + \frac{1}{16 \pi^2} \left(
2 + \log \frac{\mu^2}{M_Z^2} 
+ \frac{\widetilde{M_{N_2}}^2 \log \frac{M_Z^2}{\widetilde{M_{N_2}}^2}}{\widetilde{M_{N_2}}^2 - M_Z^2}
\right) \nonumber \\
&\quad + \frac{\widetilde{M_{D_{i3}}} \widetilde{M_{N_2}} \left(
- M_Z^2 + \widetilde{M_{N_2}}^2 \left(\log \frac{M_Z^2}{\widetilde{M_{N_2}}^2} \right)
\right)}
{32 \pi^2 \left( M_Z^2 - \widetilde{M_{N_2}}^2 \right)^2} \nonumber
\Bigg]. 
\end{align}

\begin{align}
\epsilon_{i 3}^{\phi^3 (1)} &= \frac{v}{8} \left( 
|\widetilde{Y_{13}}|^2 + |\widetilde{Y_{23}}|^2 + |\widetilde{Y_{33}}|^2 
\right) \widetilde{Y_{i3}} \\
&\times \Bigg[
\frac{\widetilde{M_{D_{i3}}}^2 \left( M_Z^4 - \widetilde{M_{N_3}}^4 - 2 M_Z^2 \widetilde{M_{N_3}}^2 \log \frac{M_Z^2}{\widetilde{M_{N_3}}^2} \right)}
{32 \pi^2 \left( M_Z^2 - \widetilde{M_{N_3}}^2 \right)^3} \nonumber \\
&\quad + \frac{1}{16 \pi^2} \left(
2 + \log \frac{\mu^2}{M_Z^2} 
+ \frac{\widetilde{M_{N_3}}^2 \log \frac{M_Z^2}{\widetilde{M_{N_3}}^2}}{\widetilde{M_{N_3}}^2 - M_Z^2}
\right) \nonumber \\
&\quad + \frac{\widetilde{M_{D_{i3}}} \widetilde{M_{N_3}} \left(
- M_Z^2 + \widetilde{M_{N_3}}^2 \left(\log \frac{M_Z^2}{\widetilde{M_{N_3}}^2} \right)
\right)}
{32 \pi^2 \left( M_Z^2 - \widetilde{M_{N_3}}^2 \right)^2} \nonumber
\Bigg].
\end{align}

\begin{align}
\epsilon_{i3}^{\phi^{\pm (2)}}  &= \frac{g^2}{64 M_W^2 \pi^2} \, 
\widetilde{M_{D_{i3}}} \, m_{\ell_i}^2 \\
&\times \Bigg[
\frac{2 \widetilde{M_{D_{i3}}} m_{\ell_i} \left(
- M_W^4 + m_{\ell_i}^4 + 2 M_W^2 m_{\ell_i}^2 
\log \frac{M_W^2}{m_{\ell_i}^2}
\right)}
{\left( -M_W^2 + m_{\ell_i}^2 \right)^3} \nonumber \\
&\quad - 2 \Bigg(
1 + \frac{M_W^2 + m_{\ell_i}^2}{M_W^2 - m_{\ell_i}^2}
+ \log \frac{\mu^2}{m_{\ell_i}^2}
- \frac{M_W^2 \left( M_W^2 + m_{\ell_i}^2 \right)
\log \frac{M_W^2}{m_{\ell_i}^2}}
{\left( M_W^2 - m_{\ell_i}^2 \right)^2}
\Bigg) \nonumber \\
&\quad - \frac{\widetilde{M_{D_{i3}}}^2}{\left( M_W^2 - m_{\ell_i}^2 \right)^4}
\Bigg(
M_W^6 - m_{\ell_i}^6 
+ 3 M_W^4 m_{\ell_i}^2 
\left( 3 - 2 \log \frac{M_W^2}{m_{\ell_i}^2} \right) \nonumber \\
&\qquad\qquad
- 3 M_W^2 m_{\ell_i}^4  
\left( 3 + 2 \log \frac{M_W^2}{m_{\ell_i}^2} \right) \nonumber
\Bigg)
\Bigg].
\end{align}

\begin{align}
\epsilon_{i3}^{\phi^{\pm}(1)} &= - \frac{g^2}{128 \, M_W^2 \, \pi^2 \, \widetilde{M_{N_3}}^3} \,
\widetilde{M_{D_{i3}}} \, m_{\ell_i}^2 \\
&\times \Bigg[
4 m_{\ell_i} \widetilde{M_{N_3}}^2 \left(
2 + \log \frac{\mu^2}{m_{\ell_i}^2}
- \frac{M_W^2 \log \frac{M_W^2}{m_{\ell_i}^2}}
{M_W^2 - m_{\ell_i}^2}
\right) \nonumber \\
&\quad + \widetilde{M_{D_{i3}}} \widetilde{M_{N_3}} \Bigg(
\frac{-4 M_W^2 m_{\ell_i} + 4 m_{\ell_i}^3 + M_W^2 \widetilde{M_{N_3}} - 3 m_{\ell_i}^2 \widetilde{M_{N_3}}}
{M_W^2 - m_{\ell_i}^2}  \nonumber \\
&\qquad + 2 (-2 m_{\ell_i} + \widetilde{M_{N_3}}) \left(
1 + \log \frac{\mu^2}{m_{\ell_i}^2}
\right) \nonumber \\
&\qquad + \frac{2 M_W^2 \left(
2 M_W^2 m_{\ell_i} - 2 m_{\ell_i}^3 - M_W^2 \widetilde{M_{N_3}} + 2 m_{\ell_i}^2 \widetilde{M_{N_3}}
\right) \log \frac{M_W^2}{m_{\ell_i}^2}}
{(M_W^2 - m_{\ell_i}^2)^2}
\Bigg) \nonumber \\
&\quad + \widetilde{M_{D_{i3}}}^2 \Bigg[
\frac{
4 m_{\ell_i}^5 - M_W^4 \widetilde{M_{N_3}}
+ 4 M_W^2 m_{\ell_i}^2 \widetilde{M_{N_3}}
- 3 m_{\ell_i}^4 \widetilde{M_{N_3}}
+ 2 M_W^2 m_{\ell_i} (2 M_W^2 + \widetilde{M_{N_3}}^2)
+ m_{\ell_i}^3 (-8 M_W^2 + 2 \widetilde{M_{N_3}}^2)
}{(M_W^2 - m_{\ell_i}^2)^2} \nonumber \\
&\qquad + 2 (2 m_{\ell_i} - \widetilde{M_{N_3}}) \left(
1 + \log \frac{\mu^2}{m_{\ell_i}^2}
\right) \nonumber \\
&\qquad - \frac{2 M_W^2 \Big(
2 M_W^4 m_{\ell_i} + 2 m_{\ell_i}^5 - M_W^4 \widetilde{M_{N_3}}
+ 3 M_W^2 m_{\ell_i}^2 \widetilde{M_{N_3}} \nonumber
- 2 m_{\ell_i}^4 \widetilde{M_{N_3}}
+ m_{\ell_i}^3 (-4 M_W^2 + 2 \widetilde{M_{N_3}}^2)
\Big) \log \frac{M_W^2}{m_{\ell_i}^2}}
{(M_W^2 - m_{\ell_i}^2)^3} \nonumber
\Bigg]
\Bigg].
\end{align}

\begin{align}
\epsilon_{i2}^{\phi^{\pm}(1)} &= \frac{g^2}{128 \, M_W^2 \, \pi^2 \, \widetilde{M_{N_2}}^3} \,
\widetilde{M_{D_{i2}}} \, m_{\ell_i}^2 \\
&\times \Bigg[
4 m_{\ell_i} \widetilde{M_{N_2}}^2 \left(
2 + \log \frac{\mu^2}{m_{\ell_i}^2}
- \frac{M_W^2 \log \frac{M_W^2}{m_{\ell_i}^2}}
{M_W^2 - m_{\ell_i}^2}
\right) \nonumber \\
&\quad + \widetilde{M_{D_{i2}}} \widetilde{M_{N_2}} \Bigg(
\frac{-4 M_W^2 m_{\ell_i} + 4 m_{\ell_i}^3 + M_W^2 \widetilde{M_{N_2}} - 3 m_{\ell_i}^2 \widetilde{M_{N_2}}}
{M_W^2 - m_{\ell_i}^2} \nonumber \\
&\qquad + 2 (-2 m_{\ell_i} + \widetilde{M_{N_2}}) \left(
1 + \log \frac{\mu^2}{m_{\ell_i}^2}
\right) \nonumber \\
&\qquad + \frac{2 M_W^2 \left(
2 M_W^2 m_{\ell_i} - 2 m_{\ell_i}^3 - M_W^2 \widetilde{M_{N_2}} + 2 m_{\ell_i}^2 \widetilde{M_{N_2}}
\right) \log \frac{M_W^2}{m_{\ell_i}^2}}
{(M_W^2 - m_{\ell_i}^2)^2}
\Bigg) \nonumber \\
&\quad + \widetilde{M_{D_{i2}}}^2 \Bigg[
\frac{
4 m_{\ell_i}^5 - M_W^4 \widetilde{M_{N_2}}
+ 4 M_W^2 m_{\ell_i}^2 \widetilde{M_{N_2}}
- 3 m_{\ell_i}^4 \widetilde{M_{N_2}}
+ 2 M_W^2 m_{\ell_i} (2 M_W^2 + \widetilde{M_{N_2}}^2)
+ m_{\ell_i}^3 (-8 M_W^2 + 2 \widetilde{M_{N_2}}^2)
}{(M_W^2 - m_{\ell_i}^2)^2} \nonumber \\
&\qquad + 2 (2 m_{\ell_i} - \widetilde{M_{N_2}}) \left(
1 + \log \frac{\mu^2}{m_{\ell_i}^2}
\right) \nonumber \\
&\qquad - \frac{2 M_W^2 \Big(
2 M_W^4 m_{\ell_i} + 2 m_{\ell_i}^5 - M_W^4 \widetilde{M_{N_2}}
+ 3 M_W^2 m_{\ell_i}^2 \widetilde{M_{N_2}} \nonumber
- 2 m_{\ell_i}^4 \widetilde{M_{N_2}}
+ m_{\ell_i}^3 (-4 M_W^2 + 2 \widetilde{M_{N_2}}^2)
\Big) \log \frac{M_W^2}{m_{\ell_i}^2}}
{(M_W^2 - m_{\ell_i}^2)^3} \nonumber
\Bigg]
\Bigg].
\end{align}

\begin{align}
\epsilon_{i2}^{\phi^{\pm}(2)}&= \frac{g^2}{64 M_W^2 \pi^2} \,
\widetilde{M_{D_{i2}}} \, m_{\ell_i}^2 \\
&\times \Bigg[
\frac{2 \widetilde{M_{D_{i2}}} m_{\ell_i} \left(
- M_W^4 + m_{\ell_i}^4 + 2 M_W^2 m_{\ell_i}^2 
\log \frac{M_W^2}{m_{\ell_i}^2}
\right)}
{(M_W^2 - m_{\ell_i}^2)^3} \nonumber \\
&\quad - 2 \Bigg(
1 + \frac{M_W^2 + m_{\ell_i}^2}{M_W^2 - m_{\ell_i}^2}
+ \log \frac{\mu^2}{m_{\ell_i}^2}
- \frac{M_W^2 (M_W^2 + m_{\ell_i}^2)
\log \frac{M_W^2}{m_{\ell_i}^2}}
{(M_W^2 - m_{\ell_i}^2)^2}
\Bigg) \nonumber \\
&\quad - \frac{\widetilde{M_{D_{i2}}}^2}{(M_W^2 - m_{\ell_i}^2)^4}
\Bigg(
M_W^6 - m_{\ell_i}^6 
+ 3 M_W^4 m_{\ell_i}^2 
\left( 3 - 2 \log \frac{M_W^2}{m_{\ell_i}^2} \right)  \nonumber \\
&\qquad\qquad
- 3 M_W^2 m_{\ell_i}^4 
\left( 3 + 2 \log \frac{M_W^2}{m_{\ell_i}^2} \right) \nonumber
\Bigg)
\Bigg].
\end{align}

\begin{align}
\epsilon_{i2}^{\phi^{3}(1)} &= \frac{v}{8} \left(
\widetilde{Y_{13}}^* \widetilde{Y_{12}} + \widetilde{Y_{23}}^* \widetilde{Y_{22}} + \widetilde{Y_{33}}^* \widetilde{Y_{32}}
\right) \widetilde{Y_{i3}} \\
&\times \Bigg[
\frac{\widetilde{M_{D_{i2}}}^2 \left(
M_Z^4 - \widetilde{M_{N_3}}^4 - 2 M_Z^2 \widetilde{M_{N_3}}^2 \log \frac{M_Z^2}{\widetilde{M_{N_3}}^2}
\right)}
{32 \pi^2 (M_Z^2 - \widetilde{M_{N_3}}^2)^3} \nonumber \\
&\quad + \frac{1}{16 \pi^2} \left(
2 + \log \frac{\mu^2}{M_Z^2}
+ \frac{\widetilde{M_{N_3}}^2 \log \frac{M_Z^2}{\widetilde{M_{N_3}}^2}}
{\widetilde{M_{N_3}}^2 - M_Z^2}
\right) \nonumber \\
&\quad + \frac{\widetilde{M_{D_{i2}}} \widetilde{M_{N_3}} \left(
- M_Z^2 + \widetilde{M_{N_3}}^2 \left(1 + \log \frac{M_Z^2}{\widetilde{M_{N_3}}^2} \right)
\right)}
{32 \pi^2 (M_Z^2 - \widetilde{M_{N_3}}^2)^2} \nonumber
\Bigg].
\end{align}

\begin{align}
\epsilon_{i2}^{\phi^3 (2)} &= \frac{v}{8} \left(
|\widetilde{Y_{12}}|^2 + |\widetilde{Y_{22}}|^2 + |\widetilde{Y_{32}}|^2
\right) \widetilde{Y_{i2}} \\
&\times \Bigg[
\frac{\widetilde{M_{D_{i2}}}^2 \left(
M_Z^4 - \widetilde{M_{N_2}}^4 - 2 M_Z^2 \widetilde{M_{N_2}}^2 \log \frac{M_Z^2}{\widetilde{M_{N_2}}^2}
\right)}
{32 \pi^2 (M_Z^2 - \widetilde{M_{N_2}}^2)^3} \nonumber \\
&\quad + \frac{1}{16 \pi^2} \left(
2 + \log \frac{\mu^2}{M_Z^2}
+ \frac{\widetilde{M_{N_2}}^2 \log \frac{M_Z^2}{\widetilde{M_{N_2}}^2}}
{\widetilde{M_{N_2}}^2 - M_Z^2}
\right) \nonumber \\
&\quad + \frac{\widetilde{M_{D_{i2}}} \widetilde{M_{N_2}} \left(
- M_Z^2 + \widetilde{M_{N_2}}^2 \left(1 + \log \frac{M_Z^2}{\widetilde{M_{N_2}}^2} \right)
\right)}
{32 \pi^2 (M_Z^2 - \widetilde{M_{N_2}}^2)^2} \nonumber
\Bigg].
\end{align}

\hspace*{\fill}

\section*{References}
\bibliographystyle{apsrev} 
\bibliography{references} 
\end{document}